\documentclass{aa}

\usepackage{graphicx,txfonts,color}
\usepackage{longtable,lscape}
\usepackage{multirow}
\usepackage{natbib}
\usepackage{xspace}
\bibpunct{(}{)}{;}{a}{}{,}

\newcommand{\kms}{km\,s$^{-1}$\xspace}
\newcommand{\mic}{$\mu$m\xspace}

\newcommand{\gastronoom}{\emph{GASTRoNOoM}\xspace}
\newcommand{\mcmax}{\emph{MCMax}\xspace}

\newcommand{\corat}{$n_\mathrm{C}/n_\mathrm{O}$\xspace}
\newcommand{\water}{H$_2$O\xspace}
\newcommand{\coabun}{$n_{\mathrm{CO}}/n_{\mathrm{H}_2}$\xspace}
\newcommand{\waterabun}{$n_{\mathrm{H}_2\mathrm{O}}/n_{\mathrm{H}_2}$\xspace}
\newcommand{\waterabuncrit}{$n_{\mathrm{H}_2\mathrm{O},\mathrm{crit}}/n_{\mathrm{H}_2}$\xspace}
\newcommand{\opr}{$OPR$\xspace}
\newcommand{\rig}{R_\mathrm{i,g}}

\newcommand{\rog}{R_\mathrm{o,g}}
\newcommand{\rid}{R_\mathrm{i,d}}
\newcommand{\rod}{R_\mathrm{o,d}}
\newcommand{\rstar}{R$_\star$\xspace}
\newcommand{\mg}{\dot{M}_\mathrm{g}}
\newcommand{\mgh}{\dot{M}_\mathrm{g,h}}
\newcommand{\mgl}{\dot{M}_\mathrm{g,l}}
\newcommand{\md}{\dot{M}_\mathrm{d}}
\newcommand{\vs}{\varv_\mathrm{stoch}}
\newcommand{\vg}{\varv_{\infty\mathrm{,g}}}
\newcommand{\vd}{\varv_{\infty\mathrm{,d}}}
\newcommand{\vlsr}{$\varv_\mathrm{LSR}$\xspace}
\newcommand{\co}{$^{12}$CO\xspace}

\newcommand{\msunyr}{\mathrm{M}_\odot\ \mathrm{yr}^{-1}}
\newcommand{\watericecoldens}{$N_{\mathrm{H}_2\mathrm{O-ice}}$\xspace}

\newcommand{\psimom}{$\psi_\mathrm{mom}$\xspace}
\newcommand{\psiemp}{$\psi_\mathrm{dens}$\xspace}
\newcommand{\psiwater}{$\psi_{\mathrm{H}_2\mathrm{O}}$\xspace}
\newcommand{\psiwaterfo}{$\psi_{\mathrm{H}_2\mathrm{O} - \mathrm{fo}}$\xspace}

\newcommand{\oh}{OH 127.8+0.0\xspace}

\begin{document}


\title{\water vapor excitation in dusty AGB envelopes\thanks{Herschel is an ESA space observatory with science instruments provided by European-led Principal Investigator consortia and with important participation from NASA.
}}
\subtitle{A PACS view of \oh}

\author{R.~Lombaert \inst{1}
\and L.~Decin \inst{1,2}
\and A.~de Koter \inst{1,2,3}
\and J.A.D.L.~Blommaert \inst{1,4}
\and P.~Royer \inst{1}
\and E.~De Beck \inst{1,5}
\and B.L.~de Vries \inst{1}
\and T.~Khouri \inst{2}
\and M.~Min \inst{2,3}}

\offprints{R. Lombaert, robinl@ster.kuleuven.be}

\institute{Instituut voor Sterrenkunde, KU Leuven, Celestijnenlaan 200D B-2401, 3001 Leuven, Belgium 
\and Astronomical Institute ``Anton Pannekoek'', University of Amsterdam, P.O.~Box 94249, 1090 GE Amsterdam, The Netherlands
\and Astronomical Institute Utrecht, University Utrecht, P.O. Box 80000, 3508 TA Utrecht, The Netherlands
\and Department of Physics and Astrophysics, Vrije Universiteit Brussel, Pleinlaan 2, 1050 Brussels, Belgium
\and Max Planck Institut f\"ur Radioastronomie, Auf dem H\"ugel 69, D-53121 Bonn, Germany}

\date{Received  / Accepted}

\authorrunning{R. Lombaert et al.}
\titlerunning{\water vapor excitation in dusty AGB envelopes}

\abstract
{AGB stars lose a large percentage of their mass in a dust-driven wind. This creates a circumstellar envelope, which can be studied through thermal dust emission and molecular emission lines. In the case of high mass-loss rates, this study is complicated by the high optical depths and the intricate coupling between gas and dust radiative transfer characteristics. An important aspect of the physics of gas-dust interactions is the strong influence of dust on the excitation of several molecules, including \water.}
{The dust and gas content of the envelope surrounding the high mass-loss rate OH/IR star \oh, as traced by Herschel observations, is studied, with a focus on the \water content and the dust-to-gas ratio. We report detecting a large number of \water vapor emission lines up to $J=9$ in the Herschel data, for which we present the measured line strengths.}
{The treatments of both gas and dust species are combined using two numerical radiative transfer codes. The method is illustrated for both low and high mass-loss-rate sources. Specifically, we discuss different ways of assessing the dust-to-gas ratio: 1) from the dust thermal emission spectrum and the CO molecular gas line strengths, 2) from the momentum transfer from dust to gas and the measured gas terminal velocity, and 3) from the determination of the required amount of dust to reproduce \water lines for a given \water vapor abundance. These three diagnostics probe different zones of the outflow, for the first time allowing an investigation of a possible radial dependence of the dust-to-gas ratio.
}
{We modeled the infrared continuum and the CO and \water emission lines in \oh simultaneously. We find a dust-mass-loss rate of $(0.5 \pm 0.1)\times 10^{-6}\ \msunyr$ and an \water ice fraction of $16\% \pm 2\%$ with a crystalline-to-amorphous ratio of $0.8\pm0.2$. The gas temperature structure is modeled with a power law, leading to a constant gas-mass-loss rate between $2\times10^{-5}\ \msunyr$ and $1\times10^{-4}\ \msunyr$, depending on the temperature profile. In addition, a change in mass-loss rate is needed to explain the $J=1-0$ and $J=2-1$ CO lines formed in the outer wind, where the older mass-loss rate is estimated to be $1 \times 10^{-7}\ \msunyr$. The dust-to-gas ratio found with method 1) is 0.01, accurate to within a {factor of three}; method 2) yields a lower limit of 0.0005; and method 3) results in an upper limit of 0.005. The \water ice fraction leads to a minimum required \water vapor abundance with respect to H$_2$ of $(1.7\pm0.2)\times 10^{-4}$. Finally, we report detecting 1612 MHz OH maser pumping channels in the far-infrared at 79.1, 98.7, and 162.9 \mic.
}
{Abundance predictions for a stellar atmosphere in local thermodynamic equilibrium yield a twice higher \water vapor abundance ($\sim$$3 \times 10^{-4}$), suggesting a 50 \% freeze-out. This is considerably higher than current freeze-out predictions. Regarding the dust-to-gas ratio, methods 2) and 3) probe a deeper part of the envelope, while method 1) is sensitive to the outermost regions. The latter diagnostic yields a significantly higher dust-to-gas ratio than do the two other probes. We offer several potential explanations for this behavior: a  clumpy outflow, a variable mass loss, or a continued dust growth.
}


\maketitle


\section{Introduction}\label{sect:intro}
Stars ascending the asymptotic giant branch (AGB) are cool and luminous, and they show pulsations with large periods and amplitudes. Their low effective temperature allows molecules and dust particles to form, with the dust playing an important role in driving the stellar wind these stars exhibit \citep{kwo1975}. As such, AGB stars are important galactic factories of interstellar gas and dust, contributing significantly to the interstellar mass budget (\citeauthor{whi1992}~\citeyear{whi1992}; \citeauthor{tie2005}~\citeyear{tie2005}). More than 70 molecular species have thus far been detected in AGB stars \citep{olo2008}. Of these, carbon monoxide (CO) is one of the most abundant circumstellar molecules after molecular hydrogen (H$_2$), locking up either all carbon atoms or all oxygen atoms, whichever is the least abundant. When carbon is more abundant (i.e.~the carbon-to-oxygen number abundance ratio \corat $ > 1$; defining C-type stars), the molecules and dust species will typically be carbon-based. When oxygen is more abundant (i.e.~\corat $ < 1$; M-type), the circumstellar envelope (CSE) will consist mainly of oxygen-based molecules and dust species \citep{rus1934,gil1969,bec1992}.

As the star ascends the AGB, the mass loss increases gradually, eventually leading to the final phase, which is suggested to be a superwind \citep{ren1981}. If the AGB star has not yet transitioned into a C-type star when it reaches the superwind phase, it is generally known as an OH/IR star, a name that stems from the presence of strong hydroxyl (OH) masers and infrared (IR) dust emission. For OH/IR stars, the comparison of mass-loss rates determined from the emission of low-excitation CO rotational transitions and those determined from the IR continuum emission appear to indicate surprisingly high dust-to-gas ratios $> 0.01$ (\citeauthor{hes1990}~\citeyear{hes1990}; \citeauthor{jus1992}~\citeyear{jus1992}; \citeauthor{del1997}~\citeyear{del1997}). As IR dust emission originates in regions closer to the stellar surface than low-excitation CO emission, therefore tracing a more recent history of the mass-loss behavior, these high dust-to-gas ratios may be spurious and in reality be a manifestation of the recent onset of a superwind (\citeauthor{jus1992}~\citeyear{jus1992}; \citeauthor{del1997}~\citeyear{del1997}).

Water (\water) vapor has been detected in CSEs of all chemical types, albeit with significantly higher abundances with respect to H$_2$ in M-type AGB stars (\waterabun $ \sim 10^{-4}$; \citeauthor{che2006}~\citeyear{che2006}; \citeauthor{mae2008}~\citeyear{mae2008}; \citeauthor{dec2010b}~\citeyear{dec2010b}). 
In these stars, \water vapor plays an important role in the energy balance because it is one of the dominant coolants in the innermost regions of the envelope thanks to its large number of far-IR transitions \citep{tru1999}. It is, however, difficult to determine \water vapor abundances accurately from \water vapor emission, owing to, e.g., a complex ro-vibrational molecular structure, multiple excitation mechanisms, and saturation effects (\citeauthor{mae2008}~\citeyear{mae2008}, \citeyear{mae2009a}; \citeauthor{dec2010b}~\citeyear{dec2010b}). 

Hitherto, a lack of \water observations {has been} hampering a detailed analysis of the \water excitation and abundance. Some \water masers and  vibrationally excited \water lines have been detected from the ground (\citeauthor{men1989}~\citeyear{men1989}; \citeauthor{men2006}~\citeyear{men2006}; see \citeauthor{mae2008}~\citeyear{mae2008} for a summary). A detailed survey of multiple \water vapor emission lines, however, requires observations made from space. Until recently, only a few space missions have detected circumstellar \water emission in the far-IR. The \emph{Infrared Space Observatory} (ISO, \citeauthor{kes1996}~\citeyear{kes1996}) found a rich \water spectrum for multiple objects, though the spectral resolution was too low to detect anything but the strongest emission lines {\citep{tru1999,bar1996,neu1996}}. 

The recently launched \emph{Herschel Space Observatory} \citep{pil2010}, allows for a breakthrough in the study of \water in AGB sources. \oh is the first high mass-loss OH/IR star observed with the \emph{Photodetecting Array Camera and Spectrometer} (PACS, \citeauthor{pog2010}~\citeyear{pog2010}) onboard Herschel. High-J CO emission has also been detected in observations made by the \emph{Heterodyne Instrument for the Far-Infrared} (HIFI, \citeauthor{deg2010}~\citeyear{deg2010}). We aim for a comprehensive study of the physics of \water in \oh by introducing a combined modeling of the gaseous and the solid state components of the outflow. We determine {the gas-mass-loss rate}, the radial abundance profile of \water vapor, the location of \water-ice formation, and the \water-ice characteristics, i.e.~the ratio of amorphous to crystalline ice particles. We also address the dust-to-gas ratio using three different diagnostics. The first uses the thermal IR continuum of the dust, the second establishes the amount of dust needed to accelerate the outflow to the observed terminal gas velocity, and the third is based on the impact of dust emission on the strength of \water lines for a given \water vapor abundance. These three diagnostics probe different zones of the circumstellar envelope, for the first time allowing an investigation of a possible radial dependence of the dust-to-gas ratio.

\section{Target selection and data reduction} \label{sect:data}
\subsection{The OH/IR star \oh} \label{sect:tarsel}
\oh, also known as V669 Cas, is a high mass-loss-rate AGB star with a relatively simple geometry. VLA maser maps of this object show an almost spherical structure \citep{bow1990}. The maps hint at possible clumpiness in the gaseous component of the CSE. Estimates for the distance to this source vary from 1.8 kpc to 7 kpc, {corresponding to a} luminosity range from $6 \times 10^3$ L$_\odot$ to $2.6 \times 10^5$ L$_\odot$ (\citeauthor{her1985}~\citeyear{her1985}; \citeauthor{eng1986}~\citeyear{eng1986}; \citeauthor{bow1990}~\citeyear{bow1990}; \citeauthor{hes1990}~\citeyear{hes1990}; \citeauthor{van1990}~\citeyear{van1990}; \citeauthor{kem2002}~\citeyear{kem2002}). We follow \citet{suh2002b}, who take the pulsational phase into account while modeling the spectral energy distribution (SED). They find a luminosity of $L_{\star,\mathrm{max}} = 3.6 \times 10^{4}$ L$_\odot$ at light maximum, $L_{\star,\mathrm{min}} = 1.0 \times 10^{4}$ L$_\odot$ at light minimum, and an {average} luminosity of $L_{\star,\mathrm{avg}} = 2.7 \times 10^4$ L$_\odot$. The last agrees with the period-luminosity relations derived by \citet{whi1991}, taking the pulsational period equal to $P = 1537 \pm 17.7$ days \citep{suh2002b}. Since the IR ISO \emph{Short Wavelength Spectrometer} (SWS; \citeauthor{deg1996}~\citeyear{deg1996}) data (observed in January 1998), as well as the PACS data (January 2010) were taken at light minimum, we take $L_\star = 1.0 \times 10^4$ L$_\odot$. This value corresponds to a distance of $d_\star = 2.1$ kpc. We assume a CO abundance of \coabun$ = 2.0 \times 10^{-4}$ \citep{dec2010a}. The gas terminal velocity $\vg= 12.5$ \kms is determined well by the width of the low-excitation transitions of CO (see Fig.~\ref{fig:data}), and is used as the primary constraint on the gas velocity field. The velocity of the system with respect to the local standard of rest is $\varv_\mathrm{LSR} = -55.0$ \kms \citep{deb2010}. The stochastic velocity of the gas in the wind is taken to be $\vs = 1.5$ \kms \citep{ski1999}. The stellar and circumstellar parameters for \oh are summarized in Table~\ref{table:inputpar}.

The CSE has been modeled by several authors who report high gas-mass-loss rates of $\dot{M}_\mathrm{g} \sim 10^{-5} - 10^{-4}\ \msunyr$ (\citeauthor{net1987}~\citeyear{net1987}; \citeauthor{bow1990}~\citeyear{bow1990}; \citeauthor{jus1992}~\citeyear{jus1992}; \citeauthor{lou1993}~\citeyear{lou1993}; \citeauthor{suh2002b}~\citeyear{suh2002b}; \citeauthor{deb2010}~\citeyear{deb2010}). Owing to the high mass-loss rate, the density in the CSE is high enough for \water ice to freeze out, shown by a strong absorption band around $3.1\ \mu$m (\citeauthor{omo1990}~\citeyear{omo1990}; \citeauthor{jus1992}~\citeyear{jus1992}; \citeauthor{syl1999}~\citeyear{syl1999}).  
\begin{table}[!t]
	{
    	\renewcommand{\arraystretch}{1.2}
    	\setlength{\tabcolsep}{2pt}
    	\caption{Overview of some stellar and circumstellar parameters of \oh. The distance to the star is denoted as $d_\star$, the stellar luminosity as $L_\star$, the CO abundance as \coabun, the pulsational period as $P$, the stellar velocity with respect to the local standard of rest as \vlsr, the stochastic velocity in the outflow as $\vs$, and the gas terminal velocity as $\vg$.}\label{table:inputpar}
  	\begin{center}
   	\begin{tabular}{lrlrlrl}\hline\hline
\multicolumn{7}{c}{Input parameters}  \\\hline
$d_\star$& 2.1 &kpc&&$P$	 & 1573 &days			\\
$L_\star$& $1.0 \times 10^4$ &L$_\odot$	&&\vlsr &-55.0 &\kms\\
\coabun  & $2.0 \times 10^{-4}$&& &$\vg$& 12.5&\kms\\	
 $\vs$  & 1.5 &\kms & & &&	\\
  	\hline
   	\end{tabular}
   	\end{center}
    	}
\end{table}
\subsection{Observations and data reduction}
\subsubsection{PACS}\label{sect:pacs}

We combined three PACS observations of \oh with six \emph{Herschel observation identifiers} (obsids, 1342189956 up to 1342189961) taken in January 2010. The first observation was performed with the standard Astronomical Observing Template (AOT) for SED.
The two others were originally obtained as part of the AOT fine-tuning campaign. The corresponding
observing modes are identical to the standard one, except that alternative chopping patterns were used.
All observations were reduced with the appropriate interactive pipeline in HIPE 8.0.1, with calibration set
32. The absolute flux calibration is based on the calibration block
(i.e.~the initial part of the observation, performed on internal calibration
sources) and is accurate to about 20\%. We have rebinned the
data with an oversampling factor of 2, i.e.~a Nyquist sampling with respect to
the native instrumental resolution. Consistency checks between the
pipeline products of the observations made with the three
chopping patterns show excellent agreement, well within the
calibration uncertainty. Since OH 127.8+0.0 is a point source, the
spectrum is extracted from the central spaxel and then point-source-corrected in all bands. We list the integrated line strengths of detected emission lines in Table~\ref{table:intint} included in the appendix. The line strengths were measured by fitting a Gaussian on top of a continuum to the lines. The reported uncertainties include the fitting uncertainty and the absolute flux calibration uncertainty of 20\%. Measured line strengths are flagged as line blends if they fulfill at least one of two criteria: 1) the full width at half maximum (FWHM) of the fitted Gaussian is larger than the FWHM of the PACS spectral resolution by at least 30\%, 2) multiple \water transitions have a central wavelength within the FWHM of the fitted central wavelength of the emission line. Other molecules are not taken into account. We caution the reader that the reported line strengths not flagged as line blends may still be affected by emission from other molecules or \water transitions not included in our modeling.

\subsubsection{{HIFI}}\label{sect:hifi}
{\oh was observed with the HIFI instrument in the HIFI Single Point AOT with dual-beam switching. The observed rotationally excited lines in the vibrational groundstate include the $J = 5-4$ (obsid 1342201529, observed in July 2010) and $J= 9-8$ (obsid 1342213357, observed in January 2011) transitions. These observations were made in the framework of the SUCCESS Herschel Guaranteed Time program (Teyssier et al., in prep.). The data were retrieved from the \emph{Herschel Science Archive}\footnote{http://herschel.esac.esa.int/Science\_Archive.shtml} and reduced with the standard pipeline for HIFI observations in HIPE 8.1. {The level 2 pipeline products were then reduced further by first applying baseline subtraction, followed by the conversion to main-beam temperature with main-beam efficiencies taken from the HIFI Observers' Manual (version 2.4, section 5.5.2.4), and finally by taking the mean of vertical and horizontal polarizations.} Finally, the $J=5-4$ line was rebinned to a resolution of 1.3 \kms and the $J=9-8$ line to a resolution of 2.2 \kms. The absolute flux calibration uncertainty of HIFI is estimated to be $\sim 10\%$ according to the HIFI Observers' Manual (version 2.4, section 5.7). However, owing to the low signal-to-noise of $\sim 4-5$ in the observed lines, we adopt a more conservative calibration uncertainty of 20\%.}

\subsubsection{Ground-based data}\label{sect:dataground}
Data for several rotationally excited lines of CO in the vibrational groundstate were obtained with the ground-based \emph{James Clerk Maxwell Telescope} (JCMT) {and the ground-based 30m telescope operated by the \emph{Institut de Radioastronomie Millim\'etrique} (IRAM). The JCMT observations include the $J = 2-1$ (observed in September 2002), $J = 3-2$ (July 2000), $J = 4-3$ (April 2000) and $J= 6-5$ (November 2002) transitions. The first three JCMT transitions were published by \citet{kem2003}, and the $J=6-5$ transition was presented by \citet{deb2010}. \citet{hes1990} published the IRAM observations including the $J=1-0$ (June 1987) and $J=2-1$ (February 1988) transitions.} We refer to these publications for the technical details concerning the data reduction. In this study, the $J=4-3$ transition is not taken into account. Considering that the line formation regions of the $J=3-2$ and the $J=4-3$ lines largely overlap, one can expect consistent line-integrated fluxes for the two lines when observed with the same telescope. No emission in the $J=4-3$ observation is significantly detected, while a line-integrated flux well above the $3\sigma$ noise level of the JCMT observation is estimated from the $J=3-2$, as well as from the other observations. This discrepancy can be caused by certain model assumptions; e.g., we do not consider that the CO $J=4$ level may be depopulated by pumping via a molecule other than CO and therefore result in a significantly decreased expected $J=4-3$ emission or by an observational issue, e.g., suboptimal pointing of the telescope. The cause of the discrepancy is not clear, so that it is safer to exclude the observation from the study.

\begin{figure}[!t]
\begin{center}$
\begin{array}{cc}
\resizebox{4.5cm}{!}{\includegraphics{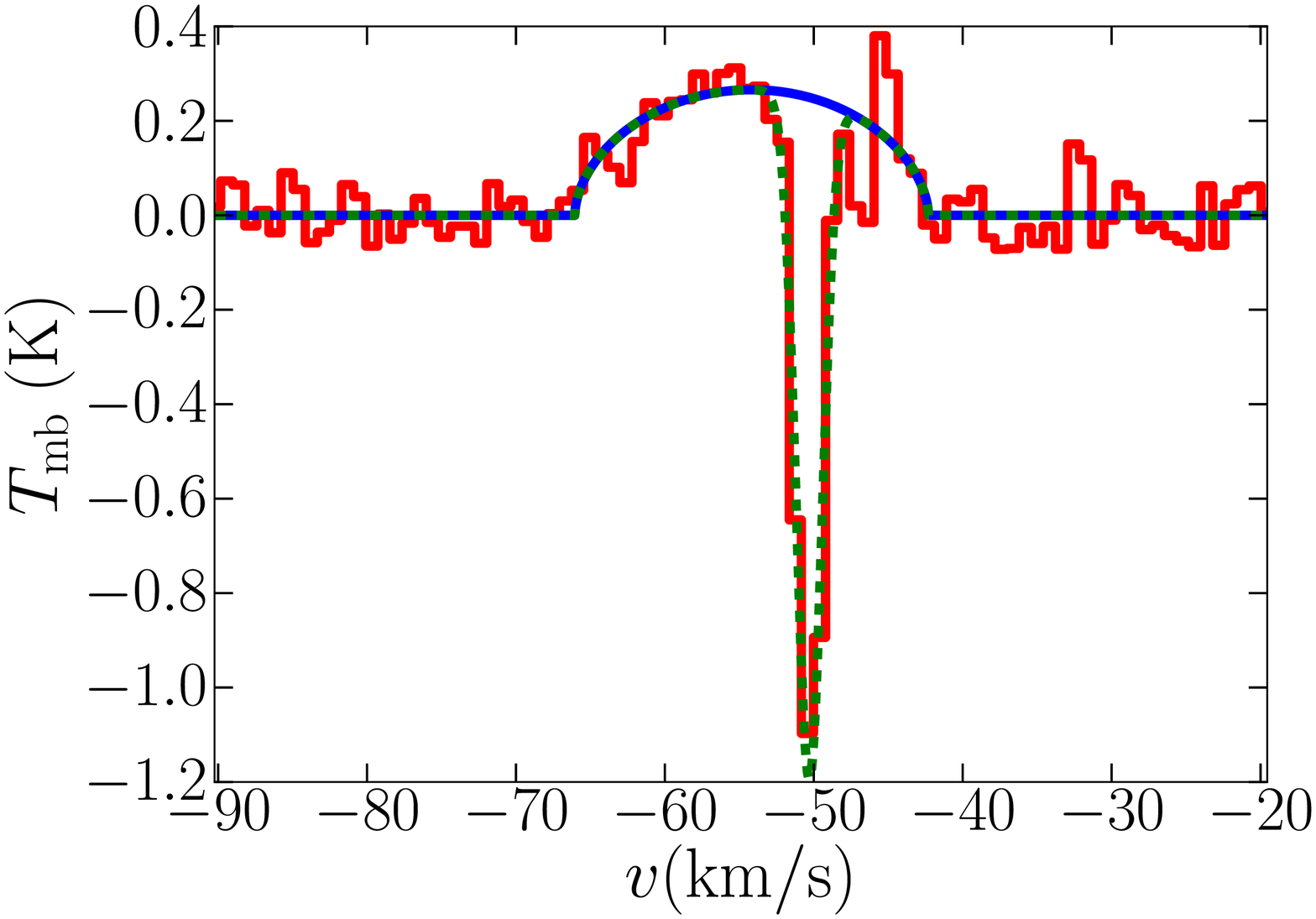}} \hspace{-0.2cm} & \resizebox{4.5cm}{!}{\includegraphics{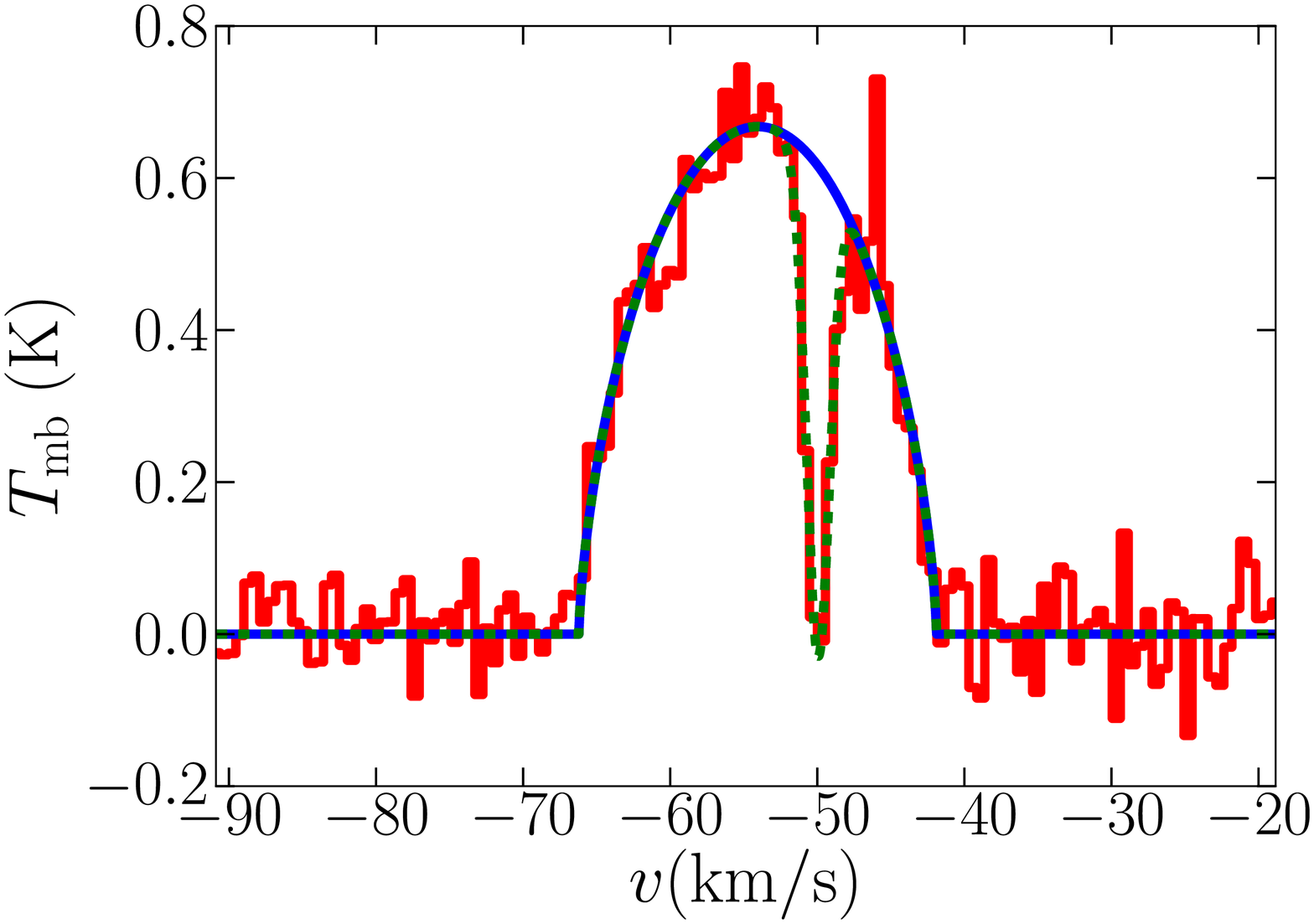}} \\
\end{array}$
\end{center}
\vspace{-0.5cm}
\caption{Ground-based JCMT observations of \oh. {The left panel shows the CO $J=2-1$ observation in red, whereas the CO $J=3-2$ is shown in the right panel. The dashed green curve gives a line profile fit including a soft-parabola and a Gaussian function. The full blue curve indicates only the soft-parabola component, which represents the emission coming from the CSE of \oh. The Gaussian component reproduces the interstellar absorption.}}
\label{fig:data}
\end{figure}

Strong CO emission at the JCMT off-source reference position contaminates the on-source $J = 2-1$ and $J=3-2$ JCMT observations after background subtraction. As shown in Fig.~\ref{fig:data}, the lines can be fitted with an analytical function equal to the sum of a soft-parabola function representing the emission profile (following \citeauthor{deb2010}~\citeyear{deb2010}) and a Gaussian function for the negative flux contribution. The Gaussian component in the fit to both observations is centered on $\sim 50$ \kms and has a width of $\sim 1$ \kms, which is a typical value for the turbulent velocity in the interstellar medium \citep{red2008}, assuming the CO emission in the off-source observation has an interstellar origin. For the CO $J=2-1$ and $3-2$ JCMT observations, we use an absolute flux calibration uncertainty of 30\% \citep{kem2003}. The CO $J=6-5$ has a low signal-to-noise ratio and is therefore treated as an upper limit with an absolute flux calibration uncertainty of 40\%. From the soft-parabola component of the $J=3-2$ observation, which both has a high signal-to-noise and suffers less from the off-source contamination than the $J=2-1$ line, we derive a gas terminal velocity $\vg \sim 12.5$ \kms. For the IRAM observations, we use the line profiles published by \citet{hes1990}, who performed a careful background subtraction to avoid an off-source CO contribution. We assume an absolute flux calibration uncertainty of 20\% for the IRAM data, taking the uncertainty involved with the background subtraction into account \citep{hes1990}.

\subsubsection{Spectral energy distribution}
The SED (see Sect.~\ref{sect:tdm}) is constructed from data obtained by the ISO-SWS and \emph{Long Wavelength Spectrometer} (LWS; \citeauthor{cle1996}~\citeyear{cle1996}; \citeauthor{swi1996}~\citeyear{swi1996}) instruments, as well as from PACS data. The ISO-SWS data were retrieved from the \citet{slo2003} database. The ISO-LWS data were taken from the ISO Data Archive\footnote{http://iso.esac.esa.int/iso/ida/} and rescaled to the calibrated ISO-SWS data. 
The ISO-LWS data are not background-subtracted, whereas the PACS data are, suggesting that more flux at long wavelengths is expected in the ISO-LWS data owing to \oh's location in the galactic plane. In addition, the PACS photometric data at 70 \mic and 160 \mic (not shown here) coincide with the PACS spectrum. The uncertainty on the absolute flux calibration of the PACS photometric data is below 15\% \citep{gro2011}. Taking these considerations into account, the ISO and PACS data agree well. The ISO-SWS and PACS data were all taken at the light minimum pulsational phase, so we assume the same stellar luminosity for both data sets and refer to the work of \citet{suh2002b} for pulsationally dependent IR continuum modeling including photometric data. Because \oh lies in the galactic plane, we corrected for interstellar reddening following the extinction law of \citet{chi2006}, with an extinction correction factor in the K-band of $A_\mathrm{K} = 0.24$ mag \citep{are1992}. 

\section{Methodology}\label{sect:methodology} 
To get a full, consistent understanding of the entire CSE, information from both gas and dust diagnostics should be coupled. Kinematical, thermodynamical, and chemical information about the circumstellar shell is derived from the molecular emission lines and the dust features by making use of two radiative transfer codes. The non-local thermodynamic equilibrium (NLTE) line radiative transfer code \gastronoom \citep{dec2006,dec2010a} calculates the velocity, temperature, and density profiles of the gas envelope, the level populations of the individual molecules, and the line profiles for the different transitions of each molecule. The continuum radiative transfer code \mcmax \citep{min2009a} calculates the dust temperature structure and the IR continuum of the envelope. These numerical codes are briefly described in Sects.~\ref{sect:gastronoom} and \ref{sect:mcmax}. In Sects.~\ref{sect:approach} to \ref{sect:dustgas}, we describe how the two codes are combined with an emphasis on the physical connections between the gaseous and dusty components. We end this section by discussing the advantage of our approach in light of molecular excitation mechanisms.
\subsection{Line radiative transfer with GASTRoNOoM}\label{sect:gastronoom}
\subsubsection{The kinematical and thermodynamical structure}\label{sect:gaskin}
The kinematical and thermodynamical structure of the CSE is calculated by solving the equations of motion of gas and dust and the energy balance simultaneously \citep{dec2006}. We assume a spherically symmetric gas density distribution. The radial gas velocity profile $\varv_\mathrm{g}(r)$ depends on the momentum transfer via collisions between gas particles and dust grains, the latter being exposed to radiation pressure from the central star. This momentum coupling is assumed to be complete \citep{kwo1975}, such that the radiative force on the dust grains can be equated to the gas drag force. 
The population of dust grains has the assumed size distribution \begin{eqnarray} n_\mathrm{d}(a,r)\ da = A(r)\ a^{-3.5}\ n_\mathrm{H}(r)\ da,\label{eq:grainsizedistr}\end{eqnarray} where $n_\mathrm{H}$ is the total hydrogen number density, $a$ the radius of the spherical dust grain, and $A(r)$ an abundance scale factor giving the number of dust particles with respect to hydrogen \citep{mat1977}. The minimum grain size considered is $a_\mathrm{min} = 0.005$ \mic and the maximum grain size $a_\mathrm{max} = 0.25$ \mic. \citet{hof2008} suggests that large grains are needed in an M-type AGB CSE to be able to drive the stellar wind through photon scattering. \citet{nor2012} have detected these large grains, with sizes up to $a \sim 0.3$ \mic, backing up our assumption of a maximum grain size of $ \sim 0.25$ \mic.

The gas kinetic temperature profile $T_\mathrm{g}(r)$ depends on the heating and cooling sources in the CSE. The heating sources taken into account are gas-grain collisional heating, photoelectric heating from dust grains, heating by cosmic rays, and heat exchange between dust and gas. The cooling modes include cooling by adiabatic expansion and the emission from rotationally excited CO and \water levels and vibrationally excited H$_2$ levels. As the difference between dust and gas velocity, the drift velocity $w(a,r)$ directly enters the equation for collisional gas heating. To calculate the contribution from the heat exchange between gas and dust, the dust-temperature profile $T_\mathrm{d}(r)$ needs to be known as well. \citet{dec2006} approximate this profile by a power law of the form 
\begin{eqnarray}
T_\mathrm{d}(r) = T_\star\left(\frac{R_\star}{2r}\right)^{2/(4+s)}, \label{eq:tdpower} 
\end{eqnarray}
where $s\approx1$ (Olofsson in \citeauthor{hab2003}~\citeyear{hab2003}). We address the dust temperature profile further in Sect.~\ref{sect:dustt}.

\subsubsection{Radiative transfer and line profiles}\label{sect:lineradtran}
The solution of the radiative transfer equations coupled to the rate equations and the calculation of the line profiles are described by \citet{dec2006}. In this work we adopt MARCS theoretical model spectra (\citeauthor{dec2007b}~\citeyear{dec2007b}; \citeauthor{gus2008}~\citeyear{gus2008}; \citeauthor{dec2010a}~\citeyear{dec2010a}) to improve the estimate of the stellar flux, as compared to a blackbody approximation. This results in more realistic absolute intensity predictions for the less abundant molecules with stronger dipole moments like \water, which are mainly excited by near-IR radiation from the central star \citep{kna1985}. {For an extensive overview of the molecular data used in this study, we refer to the appendix in \citet{dec2010a}.}

\subsection{Continuum radiative transfer with MCMax}\label{sect:mcmax}
\mcmax is a self-consistent radiative transfer code for dusty environments based on a Monte Carlo simulation \citep{min2009a}. It predicts the dust temperature stratification and the emergent IR continuum of the circumstellar envelope. We use a continuous distribution of ellipsoids (CDE, \citeauthor{boh1983}~\citeyear{boh1983}; \citeauthor{min2003}~\citeyear{min2003}) to describe the optical properties of the dust species. A CDE provides mass-extinction coefficients $\kappa_\lambda$ -- or cross-sections per unit mass -- for homogeneous particles with a constant volume, where the grain size $a_\mathrm{CDE}$ is the radius of a volume-equivalent sphere. The {CDE particle-shape model} is only valid in the Rayleigh limit, i.e. when $\lambda \gg a_\mathrm{CDE}$. For photons at wavelengths $\lambda \gg a_\mathrm{CDE}$, both inside and outside the grain, the mass-absorption coefficients $\kappa^\mathrm{a}_\mathrm{\lambda, \mathrm{CDE}}$ are independent of particle size, and the mass-scattering coefficients $\kappa^\mathrm{s}_\mathrm{\lambda, \mathrm{CDE}}$ are negligible. 

\mcmax does not include a self-consistent momentum transfer modeling procedure, i.e.~the IR continuum is calculated based on a predetermined dust density distribution $\rho_\mathrm{d}(r)$. As a standard, this density distribution is assumed to be smooth, following the equation of mass conservation  
$
\dot{M}_\mathrm{d}(r) = 4\pi\ r^2\ \varv_\mathrm{d}(r)\ \rho_\mathrm{d}(r),
$
with $\dot{M}_\mathrm{d}(r) = \md$ the dust-mass-loss rate, which is assumed to be constant, and $\varv_\mathrm{d}(r)$ the dust velocity profile, which is taken to be constant and equal to the terminal dust velocity $\vd$. Because the drift velocity is usually unknown, the dust terminal velocity is often equated to the gas terminal velocity $\vg$. In most cases, this simplification is found to be inaccurate, because the drift is nonzero \citep{kwo1975}. A possible improvement includes a customized density profile that takes a nonzero drift into account, as well as the acceleration of the dust grains derived from momentum transfer modeling (see Sect.~\ref{sect:drift}). In practice, the optical depth $\tau_\nu=1$ surface in the IR lies outside the acceleration region for high enough dust densities, so an improved density distribution in this region is not likely to affect the IR continuum of high mass-loss-rate stars. On the other hand, the effect on dust emission features in low mass-loss-rate stars may be significant.

\subsection{The five-step modeling approach}\label{sect:approach}
We solve the line radiative transfer and continuum radiative transfer using a five-step approach.
\begin{enumerate}
 \item The dust thermal IR continuum is modeled using \mcmax to obtain an initial estimate of the dust composition, dust temperature, and dust-mass-loss rate.
 \item The kinematics and thermodynamics of the gas shell are calculated with \gastronoom incorporating dust extinction efficiencies, grain temperatures, and the dust-mass-loss rate from \mcmax. This provides a model for the momentum transfer from dust to gas, hence a dust velocity profile.
 \item Given a dust-mass-loss rate, the dust velocity profile leads to a new dust-density profile for which the IR continuum model is updated. 
 \item The gas kinematical and thermodynamical structures are recalculated with the updated dust parameters.
 \item Line radiative transfer with \gastronoom is performed and line profiles are calculated.
\end{enumerate}

This five-step approach is repeated by changing various shell parameters such as the mass-loss rate and envelope sizes, until the CO molecular emission data are reproduced with sufficient accuracy. This provides a model for the thermodynamics and the kinematics of the envelope. The CO molecule is an excellent tracer for the thermodynamics of the entire gas shell because it is easily collisionally excited and relatively difficult to photodissociate.

\subsection{Incorporating gas diagnostics into the dust modeling}\label{sect:gasdust}
\subsubsection{Dust velocity profile}\label{sect:drift}
The dust velocity profile $\varv_\mathrm{d}(r)$ cannot be derived from the IR continuum emission of the dust. However, the gas terminal velocity is determined well from the width of CO emission lines observed by ground-based telescopes, providing a strong constraint on the gas kinematical model. In conjunction with the drift velocity $w(a,r)$, the gas velocity profile $\varv_\mathrm{g}(r)$ leads to $\varv_\mathrm{d}(r)$. If the momentum coupling between gas and dust is complete, one can write the drift velocity at radial distance $r$ and for grain size $a$ as (\citeauthor{kwo1975}~\citeyear{kwo1975}; \citeauthor{tru1991}~\citeyear{tru1991}; \citeauthor{dec2006}~\citeyear{dec2006})
\begin{eqnarray*}
w(a,r) &=& \varv_\mathrm{K}(a,r) \left[\sqrt{1 + x(a,r)^2}-x(a,r)\right]^{1/2},\ \mathrm{with} \\
 \varv_\mathrm{K}(a,r) &=& \sqrt{\frac{\varv_\mathrm{g}(r)}{\dot{M}_\mathrm{g}(r)c}\int Q_\lambda(a) L_\lambda d\lambda}\ ,\\
x(a,r) &=& \frac{1}{2}\left[\frac{\varv_\mathrm{T}(r)}{\varv_\mathrm{K}(a,r)}\right]^2,\ \mathrm{and}\\
\varv_\mathrm{T}(r) &=& \frac{3}{4}\left[\frac{3kT_\mathrm{g}(r)}{\mu m_\mathrm{H}}\right]^{1/2}.
\end{eqnarray*}
Here, $Q_\lambda(a)$ are the dust extinction efficiencies, $L_\lambda$ is the monochromatic stellar luminosity at wavelength $\lambda$, $\varv_\mathrm{T}(r)$ the Maxwellian velocity of the gas, $T_\mathrm{g}(r)$ the gas kinetic temperature, $k$ Boltzmann's constant, $\mu$ the mean molecular weight of the gas, and $m_\mathrm{H}$ the mass of the hydrogen atom. 

\gastronoom works with grain-size dependent extinction efficiencies, whereas we use grain-size independent CDE models for the circumstellar extinction in \mcmax. As a result, the grain-size dependent drift velocity $w(a,r)$ has to be converted to a grain-size independent average drift velocity $\bar{w}(r)$. For simplicity, we assume that the factor $\left[\sqrt{1 + x(r)^2}-x(r)\right]^{1/2}$ has a negligible effect. This assumption holds in the outer region of the CSE, where the drift velocity is much higher than the thermal velocity. The weighted drift velocity $\bar{w}(r)$ with respect to the grain-size distribution $n_\mathrm{d}(a,r)$ from Eq.~\ref{eq:grainsizedistr} can be written as
\begin{eqnarray*}
 \bar{w}(r) &=& \frac{\int \varv_\mathrm{K}(a,r)\ n_\mathrm{d}(a,r)\ da}{\int n_\mathrm{d}(a)\ da}. 
\end{eqnarray*}
Assuming a grain-size distribution between lower limit $a_\mathrm{min}$ and upper limit $a_\mathrm{max}$, this leads to
\begin{eqnarray}
\bar{w}(r) = \frac{g_\mathrm{a}\ \varv_\mathrm{K}(a_0,r)}{a_0} \label{eq:driftweight},
\end{eqnarray}
for an arbitrary grain size $a_0$ of a given drift velocity, with the weighting factor $g_\mathrm{a} = 1.25\ (a_\mathrm{max}^{-2} - a_\mathrm{min}^{-2})\times(a_\mathrm{max}^{-2.5} - a_\mathrm{min}^{-2.5})^{-1}$. 
For \gastronoom, this yields a weighting factor of $g_\mathrm{a} \simeq 0.09$. Combining $\varv_\mathrm{d}(r) = \bar{w}(r) + \varv_\mathrm{g}(r)$ with the equation of mass conservation, we find a density distribution $\rho_\mathrm{d}(r)$ that can be used in \mcmax.
\subsection{Incorporating dust diagnostics into the gas modeling}\label{sect:dustgas}
The formation of dust species in the stellar wind has a big influence on the thermal, dynamical, and radiative structure of the envelope; e.g., dust-gas collisions cause heating of the gas and drive the stellar wind, while the thermal radiation field of the dust is an important contributor to the excitation of several molecules, such as \water. An accurate description of the dust characteristics is thus paramount in any precise prediction of the molecular emission. Here, we discuss the treatment of the dust temperature, the inner shell radius, dust extinction efficiencies, and the dust-to-gas ratio. The effects of a more consistent coupling between dust and gas characteristics is described in Sect.~\ref{sect:adv}. 

\subsubsection{Dust temperature and the inner shell radius}\label{sect:dustt}
We include an average dust-temperature profile calculated with \mcmax in our gas modeling, instead of the power law in Eq.~\ref{eq:tdpower}. This average profile is calculated assuming that the dust species are in thermal contact, i.e.~distributing the absorbed radiation among all dust species such that they are at the same temperature at every radial point. We still use the independent dust temperature profiles of the different dust species - rather than the average profile - in the IR continuum modeling.

The pressure-dependent dust-condensation temperature is determined following \citet{kam2009}, setting the inner radius $\rid$ of the dust shell. Since this inner radius indicates the starting point of momentum transfer from dust to gas in the CSE, it is assumed to be the inner radius $\rig$ of the \gastronoom model as well. 


\subsubsection{Dust extinction efficiencies}\label{sect:qext}
\citet{dec2006} assume extinction efficiencies for spherical dust particles with a dust composition typical of OH/IR stars, where the main component is amorphous olivine (Mg$_\mathrm{x}$,Fe$_{1-\mathrm{x}}$)$_2$SiO$_4$ \citep{jus1992}. However, if one determines the dust composition independently by modeling the IR continuum, consistent extinction efficiencies can be derived. To convert the grain-size independent CDE mass-extinction coefficients $\kappa_\lambda$ used in \mcmax to the grain-size dependent extinction efficiencies $Q_\lambda(a)$ used in \gastronoom, the wavelength-dependent extinction coefficient $\chi_\lambda$ is written as
\begin{eqnarray*}
 \chi_\lambda = n_\mathrm{d}(a)\ \sigma_\lambda (a) = n_\mathrm{d}(a)\ Q_\lambda(a)\ \pi\ a^2,
\end{eqnarray*}
where $n_\mathrm{d}(a)$ is the number density of the dust particles in cm$^{-3}$ (see Eq.~\ref{eq:grainsizedistr}) and $\sigma_\lambda(a)$ the extinction cross-section in cm$^2$. By taking $\kappa_\lambda = \chi_\lambda\ \rho_\mathrm{d}^{-1}$, with $\rho_\mathrm{d}$ the mass density of the dust particles, it follows that
\begin{eqnarray*}
 Q_\lambda(a) = \frac{4}{3}\ \kappa_\lambda\ \rho_\mathrm{s}\ a,
\end{eqnarray*}
assuming the grains have a homogeneous grain structure. Here, $\rho_\mathrm{s}$ is the average specific density of a single dust grain. This conversion can be done as long as the Rayleigh assumption required for the CDE particle-shape model is valid for every grain size $a$ used in \gastronoom (see Sect.~\ref{sect:mcmax}). 

\subsubsection{The dust-to-gas ratio}\label{sect:d2gemp}
The dust-to-gas ratio in AGB environments is a rather ambiguous quantity and is typically assumed to be $\psi \sim 0.005 - 0.01$ (e.g.~\citeauthor{whi1994}~\citeyear{whi1994}). Different approaches can be used to estimate the dust-to-gas ratio. We assume a constant dust-to-gas ratio throughout the envelope in all of these definitions:
\begin{enumerate}
\item Models of high-resolution observations of CO emission constrain the gas-mass-loss rate $\mg$, hence the radial profile of the gas density $\rho_{\mathrm{g}}(r)$ using the equation of mass conservation. The dust-mass-loss rate $\md$ is determined from fitting the thermal IR continuum of the dust. We note that the dust velocity field used to convert $\md$ into a radial dust-density profile $\rho_{\mathrm{d}}(r)$ is obtained from the \gastronoom-modeling and accounts for drift between dust grains and gas particles. The dust-to-gas ratio is then given by $$\psi_\mathrm{dens} = \frac{\rho_\mathrm{d}}{\rho_\mathrm{g}} = \left(\frac{\md}{\vd}\right)\left(\frac{\vg}{\mg}\right).$$

\item Given the total mass-loss rate $\dot{M} = \mg + \md$, and the composition and size distribution of the dust species, \gastronoom calculates the amount of dust needed in the envelope to accelerate the wind to its gas terminal velocity $\vg$ by solving the momentum equation. This approach depends on the efficiency of the momentum coupling between the dust and gas components of the CSE. We assume complete momentum coupling, but we point out that this assumption does not always hold (\citeauthor{mac1992}~\citeyear{mac1992}; \citeauthor{dec2010a}~\citeyear{dec2010a}). The empirical value of $\vg$ is determined from high-resolution observations of low-excitation emission lines, such as CO $J=1-0$. The dust-to-gas ratio determined via this formalism will be denoted as \psimom.

\item In case of a high mass-loss rate, CO excitation is not sensitive to the dust emission, which allows one to constrain the gas kinetic temperature profile and the $\mg$-value by modeling the CO emission. In contrast, a main contributor to the excitation of \water is thermal dust emission. This allows one to determine the amount of dust required to reproduce the observed line intensities for a given \water vapor abundance. This leads to a dust-to-gas ratio denoted as \psiwater, which depends on the adopted \water vapor abundance.
\end{enumerate}

\subsection{Advantages of combined dust and gas modeling: molecular excitation}\label{sect:adv}
Calculating theoretical line profiles for molecular emission strongly depends on several pumping mechanisms to populate the different excitation levels, some of which are connected to the dust properties of the outflow. The most common mechanisms to populate the rotational levels in the vibrational groundstate include:
\begin{enumerate}
 \item Collisional excitation: Low-energy excitation is usually caused by collisions between a molecule and H$_2$. This mechanism becomes more important with higher densities due to the more frequent collisions. For instance, the ground-vibrational level of CO is easily rotationally excited (transitions at $\lambda > 200$ \mic).
 \item Excitation by the near-IR radiation field: The near-IR stellar continuum photons can vibrationally excite molecules. The vibrational de-excitation then happens to rotationally excited levels in lower vibrational states, with the rotational level being determined by quantum-mechanical selection rules. For instance, the first vibrational state ($\lambda \sim 4.2$ \mic) of CO and the $\nu_1=1$ ($\lambda \sim 2.7$ \mic), $\nu_2=1$ ($\lambda \sim 6.3$ \mic), and $\nu_3=1$ ($\lambda \sim 2.7$ \mic) vibrational states of \water are excited this way. If the dust content of a CSE is high, a significant fraction of the stellar near-IR photons are absorbed and re-emitted at longer wavelengths, and cannot be used for vibrational excitation of molecules.
 \item Excitation by the diffuse radiation field: The diffuse field is mainly the result of thermal emission by dust and the interstellar background radiation field. These photons allow rotational excitation to levels that require energies that are too high to be accessed through collisional excitation, and too low to be excited by absorption from the stellar near-IR radiation field. For instance, the ground-vibrational level of \water is rotationally excited through photons provided by the diffuse field ($\lambda \sim 10 - 200$ \mic). Increasing the dust content causes more pumping through this channel. 
\end{enumerate}
The relative importance of these mechanisms strongly depends on the Einstein coefficients and on the local physical conditions of both the dust and gas components of the CSE.

\begin{figure}[!t]
\resizebox{\hsize}{!}{\includegraphics{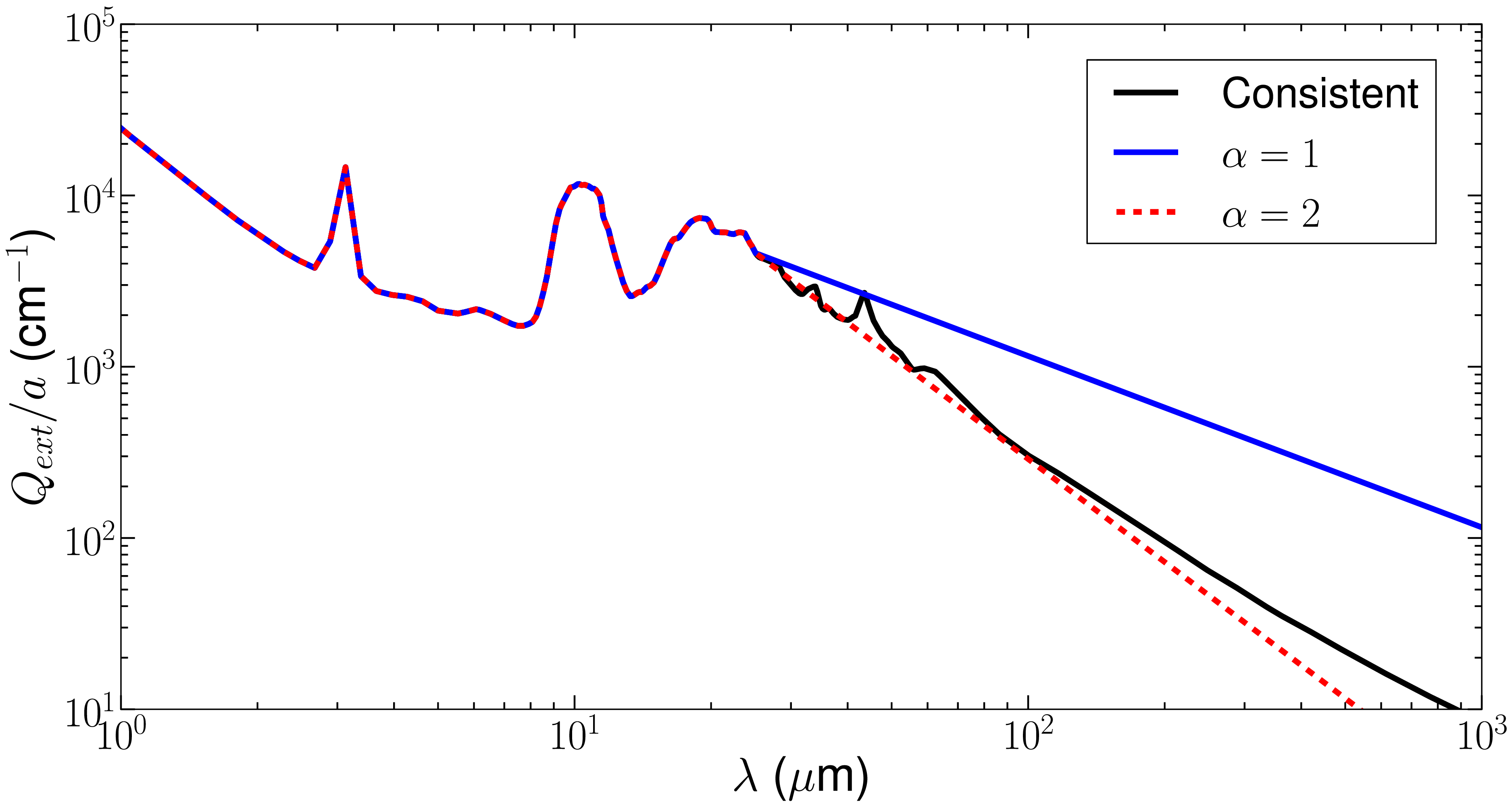}}
\caption{The dust extinction efficiencies divided by grain size (in cm$^{-1}$) versus wavelength (in $\mu$m) used for the models shown in Figs.~\ref{fig:comp_water_qext} and \ref{fig:comp_water_d2g}. At $\lambda < 25\ \mu$m the profiles are identical. From $25\ \mu$m onward, the blue full line and the red dashed line show a profile where the region at $\lambda > 25\ \mu$m is replaced with a power law of the form $\mathrm{Q}_{\mathrm{ext}}/a \sim \lambda^{-\alpha}$ assuming $\alpha = 1$ and $\alpha = 2$, respectively. The black full line is representative of a typical oxygen-rich OH/IR extinction profile as used in \mcmax, for which the dust composition is given in Table~\ref{table:dust}.}
\label{fig:qext_long_wav}
\end{figure}

To show the effect of dust on line emission predictions for a few selected lines of CO and \water in different excitation regimes accessible in the PACS wavelength domain, we use a standard input template (Table~\ref{table:inputpar}) and vary one parameter at a time. We give an overview for high ($\mg \sim 5.0\times 10^{-5}\ \msunyr$) and low ($\mg \sim 1.0\times 10^{-7}\ \msunyr$) mass-loss rates of the most significant effects including the condensation radius, the dust extinction efficiency profile, and the dust-to-gas ratio. For simplicity, we assume a power law for the gas temperature profile corresponding to Model 1 in Table~\ref{table:templaws}. The extinction efficiency profiles under consideration are shown in Fig.~\ref{fig:qext_long_wav}. We present profiles for the CO $J=16-15$ transition and the \water $2_{1,2}-1_{0,1}$ and $4_{2,3}-4_{1,4}$ transitions, all in the vibrational ground state. Figure \ref{fig:comp_water_qext} displays the high mass-loss-rate case, and Fig.~\ref{fig:comp_water_d2g} the low mass-loss-rate case. We discuss the effects below.

\subsubsection{The condensation radius}
\begin{figure}
\resizebox{\hsize}{!}{\includegraphics{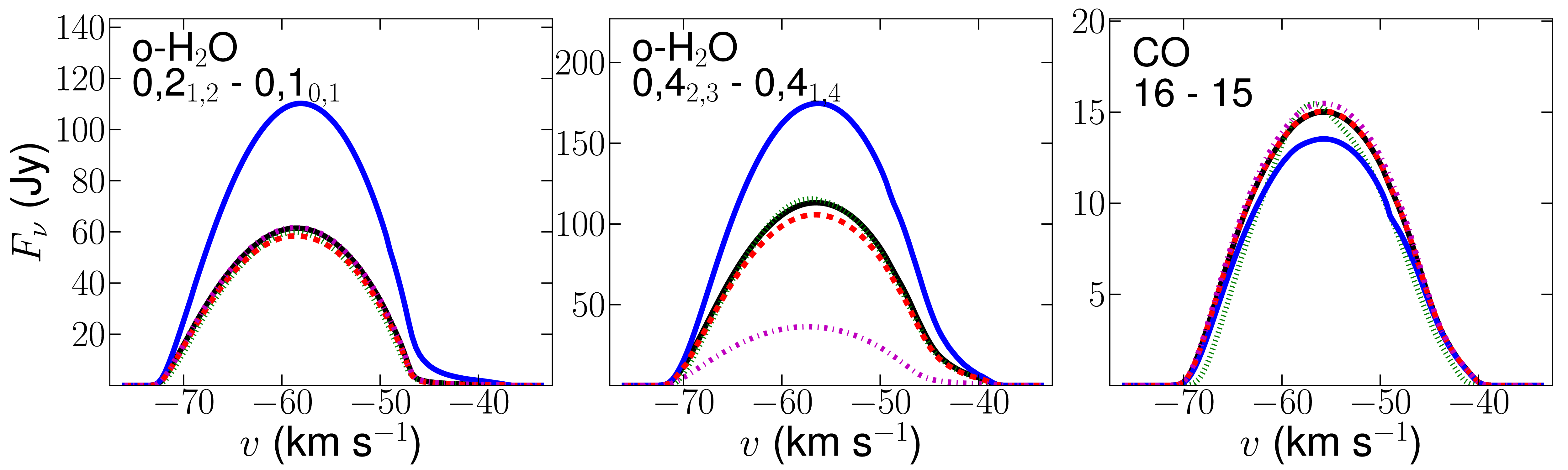}}
\caption{Line profile predictions for the high mass-loss-rate case $\mg = 5.0 \times 10^{-5}\ \msunyr$. The full black curve corresponds to the standard model with the inner radius of the gas shell $\rig = 3$ \rstar, the black extinction efficiency profile from Fig.~\ref{fig:qext_long_wav} and $\psi = 0.01$. In all other models only a single property is modified. The dotted green curve {(which coincides with the other curves)} assumes $\rig = 10$ \rstar, the full blue and dashed red curves apply the blue and red extinction efficiency profiles from Fig.~\ref{fig:qext_long_wav} and the dashed-dotted magenta curve assumes $\psi=0.001$ (see Sect.~\ref{sect:adv} for more details).}
\label{fig:comp_water_qext}
\end{figure}

In the high mass-loss-rate case, the condensation radius is not expected to have a strong influence on the theoretical line profiles thanks to the high opacity of the envelope. Indeed, the full black (condensation radius $\rig = 3$ R$_\star$) and dotted green ($\rig = 10$ R$_\star$) models coincide in Fig.~\ref{fig:comp_water_qext} and the transitions have a parabolic line profile typical for optically thick winds. The lines shown here are formed at $r > 20$ \rstar when the wind has already been fully accelerated, i.e.~farther from the stellar surface than the condensation radius used for the green model.

\begin{figure}[!t]
\resizebox{\hsize}{!}{\includegraphics{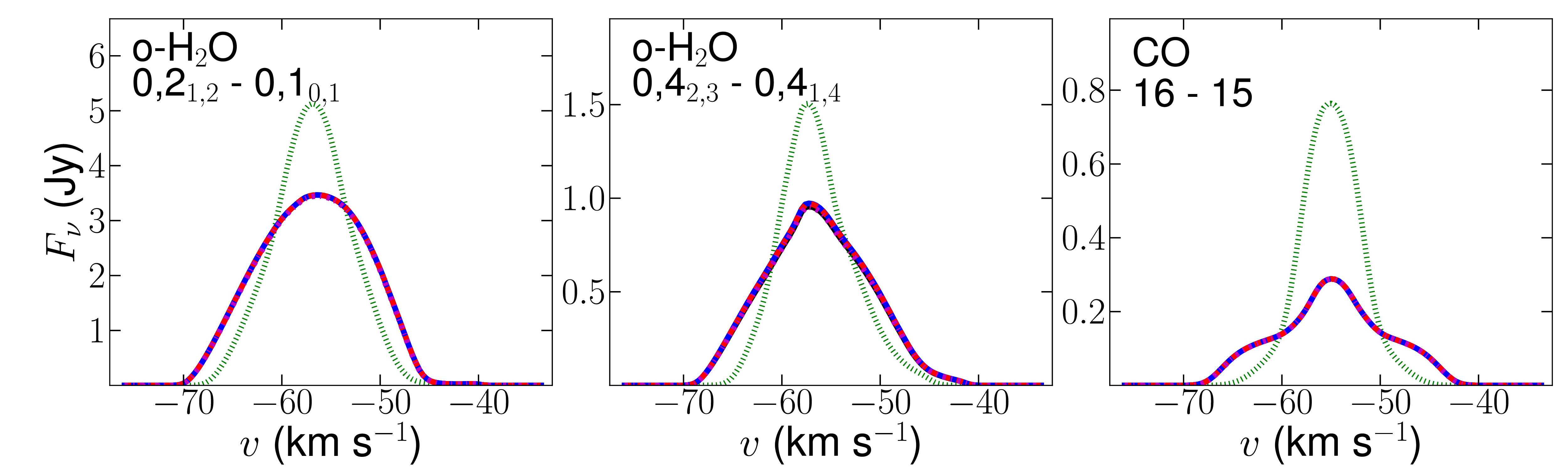}}
\caption{As Fig.~\ref{fig:comp_water_qext}, with $\mg = 1.0 \times 10^{-7}\ \msunyr$. All but the dotted green curve coincide. }
\label{fig:comp_water_d2g}
\end{figure}

In the low mass-loss-rate case, the line formation regions of the lines discussed here are located in the dust condensation region and the acceleration zone. Increasing the condensation radius in the low mass-loss rate model results in the removal of a relatively large amount of dust and effectively moves the acceleration zone outward. This manifests itself in the shape of the line profile. In the green model ($\rig = 10$ R$_\star$) in Fig.~\ref{fig:comp_water_d2g}, the line formation regions are located where the wind is accelerated. As a result, the lines exhibit a narrow Gaussian profile \citep{buj1991,dec2010a}. In the black standard model, a narrow and a broad component are visible in the CO line, indicating that the line is formed both in a region where the wind is still being accelerated, and in a region where the wind has reached its terminal gas velocity. The \water $2_{1,2}-1_{0,1}$ line, however, is only formed in the part of the wind that has just reached the terminal gas velocity and leans toward a parabolic profile typical for an optically thick wind tracing only the terminal velocity. Even though dust is unimportant for the excitation of CO, its indirect influence through the optical depth of the inner region of the envelope highlights the importance of dust formation sequences and of the stellar effective temperature, which are often poorly constrained. 

\subsubsection{The dust opacity law}
Often, the dust extinction efficiency profile is approximated by a power law, $\mathrm{Q}_{\mathrm{ext}} \sim \lambda^{-\alpha}$, especially at wavelengths $\lambda >
25$ \mic. \citet{lam1999} propose $\alpha \sim 2$, while \citet{jus1992} suggest $\alpha \sim 1$ up to 1.5. \citet{tie1987} propose to use $\alpha \sim 2$ for crystalline grains and $\alpha \sim 1$ for amorphous grains. An AGB envelope is usually dominated by amorphous material (up to at least 80 \% of the dust is amorphous, e.g.~\citeauthor{dev2010}~\citeyear{dev2010}). However, Fig.~\ref{fig:qext_long_wav} shows that $\alpha = 2$ is a better approximation of the dust extinction efficiency profile as calculated with \mcmax for \oh. 

Comparing the three theoretical profiles for the high mass-loss rate case indicates the importance of the dust extinction efficiency profiles. This is expected because these efficiencies determine the thermal emission characteristics of the grains. The relative change of an \water line depends not only on the opacity law, but also on where the line is formed in the wind and on the spectroscopic characteristics of the line; i.e., for different excitation frequencies the dust radiation field will have a different effect. It is not straightforward to predict how these changes will show up for given assumptions about the dust extinction efficiency profile. If the excitation includes channels at wavelengths $\lambda \sim 10 - 200$ \mic (i.e.~excitation mechanism 3), \water excitation is very sensitive to the properties of the dust grains in the CSE. At low mass-loss rates, however, the dust content is too low for this mechanism to contribute significantly, such that \water excitation is controlled by the stellar radiation field in the near-IR (i.e.~excitation mechanism 2).

\subsubsection{The dust-to-gas ratio}
At high mass-loss rates, the sensitivity of \water excitation to the dust properties becomes very clear when comparing the low and high dust-to-gas ratio models in Fig.~\ref{fig:comp_water_qext}. To demonstrate this sensitivity, we consider first the \water $4_{2,3}-4_{1,4}$ line. The excitation mechanism for the \water $4_{2,3}$ level involves first absorbing photons at $\lambda \sim 273$ \mic, where the dust radiation field is weak, and subsequently at $\lambda \sim 80$ \mic, where the dust radiation field dominates. Decreasing the dust-to-gas ratio implies that fewer photons are available for the channel at $\lambda \sim 80$ \mic, decreasing the population of the $4_{2,3}$ level. As a result, the strength of the \water $4_{2,3}-4_{1,4}$ emission line is decreased significantly. Populating the \water $2_{1,2}$ level, on the other hand, only involves channels at $\lambda \sim 180$ \mic, where the dust radiation field is again weak. As a result, the \water $2_{1,2}-1_{0,1}$ line is not affected by a decrease in the dust-to-gas ratio.

Both \water lines are affected by a change in the dust extinction efficiency profile. A profile with a different slope ($\alpha = 1$ as opposed to $\alpha = 2$ in this example, see Fig.~\ref{fig:qext_long_wav}) results in a relatively stronger dust radiation field at wavelengths $\lambda > 150$ \mic as compared with the dust radiation field at $\lambda \sim 80$ \mic. As a result, both \water lines are affected because the dust radiation field becomes stronger with respect to the underlying stellar and interstellar background radiation field at $\lambda > 150$ \mic. CO emission is not noticeably affected when changing the dust-to-gas ratio, indicating that collisional excitation dominates for this molecule. 

Ultimately, if collisions are not energetic enough to have a significant impact, it is the balance between 1) the dust, 2) the stellar, and 3) the interstellar background radiation fields at all wavelengths involved in populating a given excitation level that will determine the effect of different dust properties on molecular line strengths.

\section{Case study: the OH/IR star \oh} \label{sect:case}
We applied the combined modeling with \gastronoom and \mcmax to the OH/IR star \oh. Table~\ref{table:modelpar} gives the modeling results, which are discussed in this section.
\begin{table}[!t]
	{
    	\renewcommand{\arraystretch}{1.2}
    	\setlength{\tabcolsep}{2pt}
    	\caption{{Modeling results for \oh, associated with Model 2 in Table~\ref{table:templaws}.} $T_\star$ gives the stellar effective temperature; $\rid$ and $\rig$ the dust and gas inner radii respectively; $\rog$ the photodissociation radius of $^{12}$CO; $\rod$ the dust outer radius; $\md$ and $\mg$ the dust and gas mass-loss rates; $\vd$ the dust terminal velocity; \psimom, \psiemp and \psiwater the dust-to-gas ratios derived from three different methods (see Sect.~\ref{sect:d2gemp}); \watericecoldens the \water ice column density; \opr the ortho-to-para \water ratio; and \waterabuncrit the critical \water vapor abundance with respect to H$_2$.}\label{table:modelpar}
  	\begin{center}
   	\begin{tabular}{lrlrlrl}\hline\hline
\multicolumn{7}{c}{Modeling results}  \\\hline
$T_\star$ 	& 3000 &K	&&\psimom    	& $\geq 0.05 \times 10^{-2}$&\\
$\rid = \rig$	& 7.0 &\rstar 	&&\psiemp    	& $1.0 \times 10^{-2}$&\\
$\rog$	& $50 \times 10^3$ &\rstar&	&\psiwater  & $\leq 0.5 \times 10^{-2}$&\\
$\rod$	& $7 \times 10^3$ &\rstar&	&\watericecoldens & $3.9 \times 10^{17}$&cm$^{-2}$\\
$\mg$& $5 \times 10^{-5}$&$\msunyr$ & &\opr	    & 3		&	  \\
$\md$& $5 \times 10^{-7}$&$\msunyr$&&\waterabuncrit & $1.7 \times 10^{-4}$&\\
$\vd$&13.6 &	\kms	  &&&& \\
  	\hline
   	\end{tabular}
   	\end{center}
    	}
\end{table}
\subsection{Thermal dust emission}\label{sect:tdm}
To model the IR continuum of \oh, we followed the five-step approach presented in Sect.~\ref{sect:approach}. With the assumed parameters listed in Table~\ref{table:inputpar}, there are few parameters left to adapt in order to reproduce the observed IR continuum. The inner radius $\rid$ was fixed by considering pressure-dependent dust condensation temperatures. The stellar effective temperature $T_\star$ has no influence on the IR continuum of the dust due to the high optical depth of the wind of \oh and is constrained to some extent by the CO emission modeling. The dust terminal velocity $\vd$ was derived from the momentum transfer between gas and dust. 

\begin{figure}[!t]
\resizebox{\hsize}{!}{\includegraphics{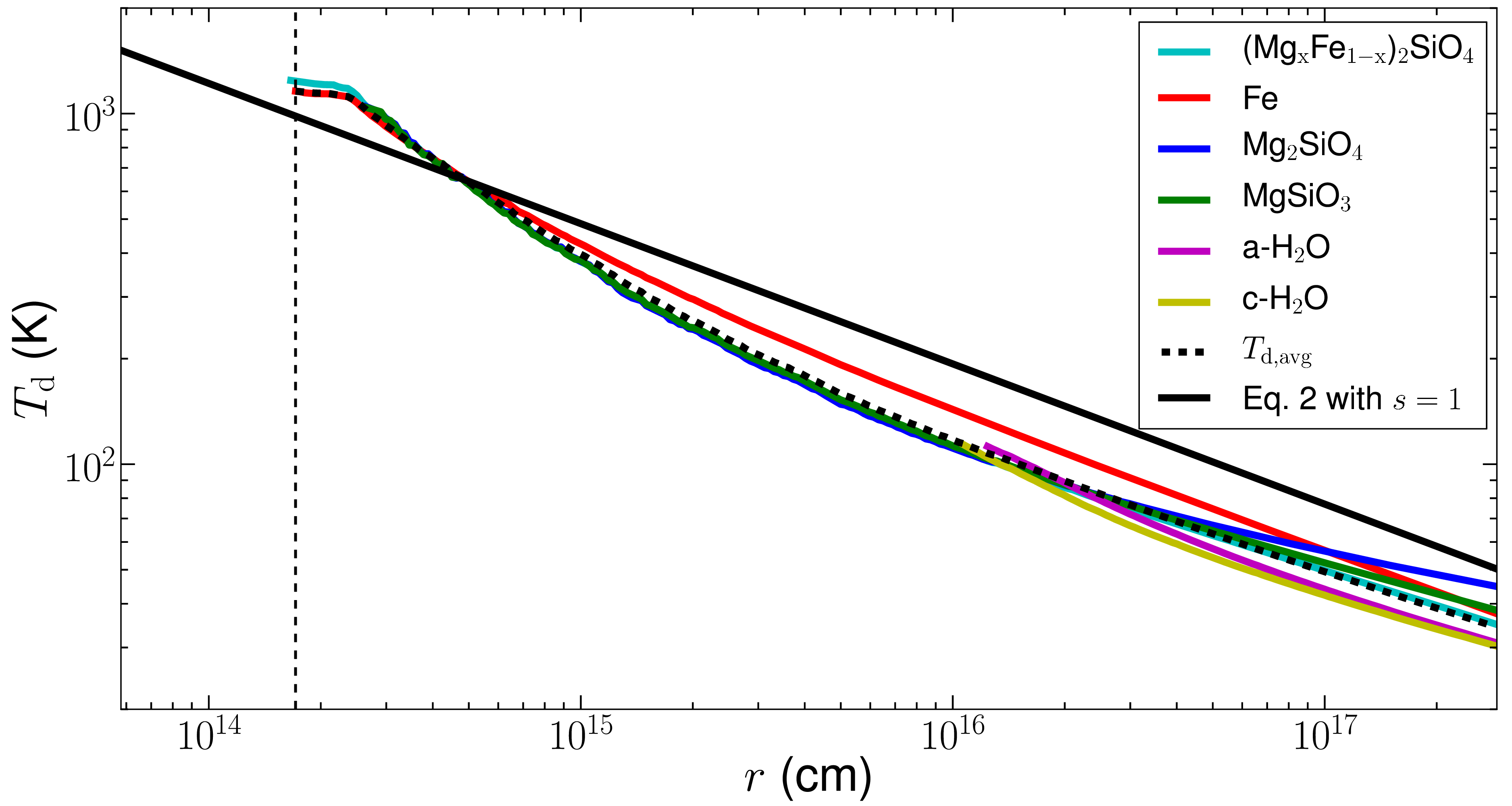}}
\caption{The dust-temperature profiles for \oh as modeled with \mcmax. The full colored lines indicate the specific dust species: cyan for amorphous silicates, red for metallic iron, blue for forsterite, green for enstatite, magenta for amorphous \water ice, and yellow for crystalline \water ice. Each of these profiles are cut off at the condensation temperature. The dashed black line gives the mean dust temperature profile. The full black line shows the power law from Eq.~\ref{eq:tdpower}, with $s=1$. The vertical dashed line indicates the inner radius of the dust shell.}
\label{fig:dusttemp}
\end{figure}

This only leaves the dust-mass-loss rate $\md$, the outer radius of the dust shell $\rod$, and the dust composition as free parameters for fitting the thermal dust emission features and the overall shape of the IR continuum. The parameter $\rod$ was chosen such that the emergent flux at long wavelengths matches the PACS data well, in agreement with the model suggested by \citet{kem2002}. \citet{syl1999} show that the spectral features in the 30 to 100 \mic range can be reproduced by a combination of amorphous silicates, forsterite, enstatite, and crystalline \water ice. Following \citet{kem2002}, metallic iron was also included. The theoretical extinction coefficients of amorphous silicate were calculated from a combination of amorphous olivines with different relative magnesium and iron fractions, determined by modeling the dust features in the IR continuum of the O-rich AGB star Mira \citep{dev2010}. The dust species and their condensation temperatures, as well as their mass fractions, are listed in Table~\ref{table:dust}. The dust-mass fractions are given in terms of mass density of the dust species with respect to the total dust-mass density, assuming all six modeled dust species have been formed. Figure \ref{fig:dusttemp} shows the temperature profiles of each dust species. Also shown is the average dust temperature profile $T_\mathrm{d,avg}${ that is adopted as }the input dust-temperature profile for \gastronoom in our five-step approach. Our results for the dust composition agree well with those of \citet{kem2002}. We find a higher forsterite abundance and slightly higher metallic iron abundance, whereas the amorphous silicate abundance is lower. These differences are minor.

\begin{figure}[!t]
\resizebox{\hsize}{!}{\includegraphics{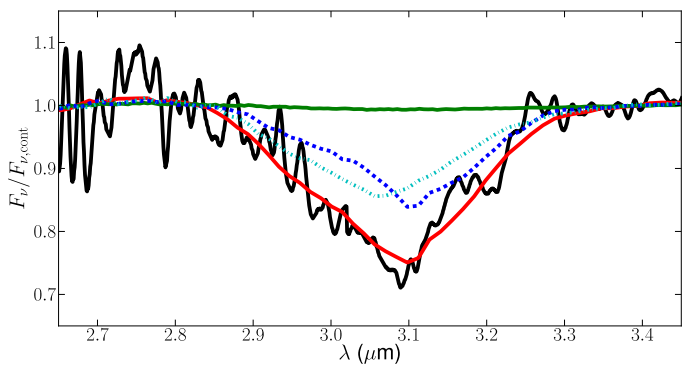}}
\caption{The 3.1 \mic ice absorption feature. The continuum-divided ISO-SWS data are shown in black. The red curve gives the best fit model and the green curve gives the model without \water ice. The dashed blue and dotted cyan curve give the contributions from crystalline and amorphous \water ice, respectively.}
\label{fig:ice}
\end{figure}

The mass fraction of crystalline and amorphous \water ice is determined by fitting the 3.1 \mic absorption feature in the continuum-divided ISO-SWS data, see Fig.~\ref{fig:ice}. The slightly shifted peak position around 3.1 \mic in the mass extinction coefficients of amorphous and crystalline ice allows one to reproduce the shape and strength of this absorption feature. We find a crystalline to amorphous \water ice ratio of $0.8\pm0.2$ and a total relative mass fraction of $(16 \pm 2)\%$ for \water ice, which leads to a radial column density of \watericecoldens$=(3.9\pm0.5) \times 10^{17}$ cm$^{-2}$.

\citet{syl1999} and \citet{kem2002} have modeled the IR continuum of \oh extensively. Using only crystalline \water ice, they find \watericecoldens$ = 5.5 \times 10^{17}$ cm$^{-2}$ and \watericecoldens$ = 8.3 \times 10^{17}$ cm$^{-2}$, respectively. \citet{dij2006} have done a theoretical study of \water ice formation \citep{dij2003} to calculate the expected \water ice mass fractions in OH/IR stars. For a CSE with parameters similar to what we find for \oh, they expect that only $2\%$ of the total dust mass is \water ice, which is a factor of 5 lower than the \citet{kem2002} results and a factor of 8 lower than our results. However, they assumed an initial \water vapor abundance of $1 \times 10^{-4}$ in their \water ice formation models, which is a rather low estimate for an OH/IR star \citep{che2006}. More \water vapor may lead to the formation of more \water ice and would be more in line with our results. Moreover, following their \water ice formation models, \citet{dij2006} show that no strong \water ice features are expected in the IR continuum at $43\ \mu$m and $62\ \mu$m because most of the \water ice is predicted to be amorphous. Unlike this theoretical result, they point to significant fractions of crystalline \water ice in the spectra of many sources, in agreement with the large crystalline fraction that we find for \oh. They suggest several explanations for this behavior, including a high mass-loss rate over luminosity ratio, axisymmetric mass loss, and clumpiness of the wind, all of which were not taken into account in their ice formation models.

\begin{table*}[!ht]
	{
    	\setlength{\tabcolsep}{5pt}
    	\caption{The dust composition of \oh's CSE. Listed are the dust species with their chemical formula, their specific density $\rho_\mathrm{s}$, the condensation temperature $T_\mathrm{cond}$, the mass fraction of the dust species (given as the mass density of the dust species with respect to the total dust-mass density $\rho_{\mathrm{species}}/\rho_\mathrm{d}$, assuming all dust species have been formed), and the reference to the optical data for the opacities. The references for the optical constants of the dust species are as follows: 1.~\citet{dev2010} and references therein; 2.~\citet{jag1998a}; 3.~\citet{ser1973}; 4.~\citet{hen1996}; 5.~\citet{war1984}; 6.~\citet{ber1969}.}\label{table:dust}
  	\begin{center}
   	\begin{tabular}[c]{lllllll}\hline\hline\rule[0mm]{0mm}{3mm}
   Dust species	 &  Chemical formula    &   $\rho_\mathrm{s}$       & $T_\mathrm{cond}$     & $\rho_{\mathrm{species}}/\rho_\mathrm{d}$	&Reference \\
                 &                 &   (g cm$^-3$)   & (K)                        &  (\%)    &		\\\hline
   Amorphous silicate  &  (Mg$_\mathrm{x}$Fe$_\mathrm{1-x}$)$_2$SiO$_4$ & 3.58 & 1100  & 69 & 1 \\
   Enstatite     &  MgSiO$_3$      &   2.80         & 950                       & 3  & 2 \\
   Forsterite    &  Mg$_2$SiO$_4$  &   3.30         & 950                       & 7  & 3 \\
   Metallic iron &  Fe             &   7.87         & 1150                       & 5  & 4 \\
   Crystalline water ice     &  c-\water         &   1.00         & 110                        & 7 & 5 \\
   Amorphous water ice     & a-\water         &   1.00         &  100                       & 9 & 5,6 \\  	\hline
   	\end{tabular}
   	\end{center}
	}
\end{table*}

\begin{figure}[!t]
\resizebox{\hsize}{!}{\includegraphics{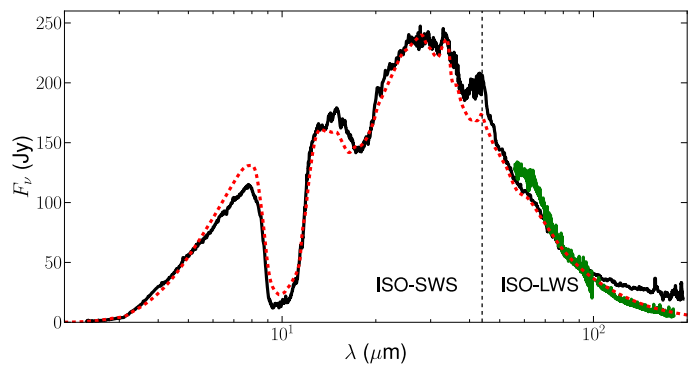}}
\caption{The SED of \oh. In black the combined ISO-SWS and LWS data are shown; in green the PACS data are given. The dashed red curve is our best-fit model. The vertical dashed black line indicates the transition between the ISO-SWS and ISO-LWS data.}
\label{fig:sed}
\end{figure}

For the dust composition described above we find a dust-mass-loss rate of $\md = (5.0 \pm 1.0) \times 10^{-7}\ \msunyr$. This agrees well with results previously obtained: $\md = 4.0 \times 10^{-7}\ \msunyr$ by \citet{suh2004} and $\md = (7\pm1) \times 10^{-7}\ \msunyr$ by \citet{kem2002}, both assuming spherical dust grains. The use of the CDE particle-shape model results in higher extinction efficiencies relative to spherical particles \citep{min2003}, in principle implying the need for less dust to fit the IR continuum of the dust. The choice of particle model does not significantly influence the relative mass fractions of the dust species. The resulting SED model, as well as the data, is shown in Fig.~\ref{fig:sed}. We lack some IR continuum flux in the region 40 \mic $ < \lambda < $ 70 \mic in our model, which is a problem that has been indicated by previous studies of OH/IR stars, e.g.~\citet{kem2002} and \citet{dev2010}.

\subsection{Molecular emission}\label{sect:molecem}
We focus here on modeling the CO and \water emission lines. Apart from these molecules, notable detections in the PACS spectrum concern OH emission at $\lambda \sim 79.1$ \mic, $\sim 98.7$ \mic and $\sim 162.9$ \mic. The line strengths of these emission lines are listed in Table~\ref{table:intint}. Because the OH emission occurs in doublets, the line strengths of both components have been summed. We refer to \citet{syl1997} for details on OH spectroscopy. These detections agree with the OH rotational cascade transitions involved in some of the far-IR pumping mechanisms suggested as being responsible for the 1612 MHz OH maser \citep{eli1976,gra2005}. Additional OH rotational cascade transitions are expected in the PACS wavelength range at $\lambda \sim 96.4$ \mic and $\sim 119.4$ \mic, but they are not detected. These results are in accordance with \citet{syl1997}, who have searched for the 1612 MHz OH maser channels in the ISO data of the yellow hypergiant IRC+10420. The three strongest emission lines were found at the same wavelengths as our OH detections, while the two other rotational cascade lines in the PACS wavelength range were significantly weaker, if detected at all. To our knowledge, this is the first detection of the 1612 MHz OH maser formation channels in the far-infrared in an AGB CSE. Owing to the complexity of maser formation and the spectroscopy of OH, however, we do not include these OH emission lines in the analysis.

\subsubsection{CO emission}\label{sect:coemission}
{We assume that the dust-to-gas momentum transfer initiates the stellar wind at the inner radius $\rid$ of the dust shell derived from the pressure-dependent dust condensation temperatures (see Sect.~\ref{sect:tdm}).} The outer radius $\rog$ of the gas shell is taken as equal to the photodissociation radius of CO, following the formalism of \citet{mam1988}. This leaves the gas-mass-loss rate $\mg$, the stellar effective temperature $T_\star$, and the gas kinetic temperature profile $T_\mathrm{g}(r)$ as free parameters to model the CO emission lines. 

In the five-step approach, the thermodynamics of the gas shell can be calculated consistently for steps 2 and 4. If the \water vapor abundance is high (\waterabun $ > 10^{-6}$), \water cooling becomes one of the dominant processes in the gas thermodynamics \citep{dec2006}. This introduces a significant uncertainty in the gas-temperature profile if the \water vapor abundance is not well constrained. We therefore opt to parametrize the temperature structure. Using a grid calculation for several temperature structures and for a wide range of gas-mass-loss rates, we constrain $T_\mathrm{g}(r)$ empirically for \oh. The grid probes five free parameters: the gas-mass-loss rate, ranging from $1.0 \times 10^{-5}\ \msunyr$ to $2.0 \times 10^{-4}\ \msunyr$; the stellar effective temperature, ranging from 2000 K to 3500 K; and the gas kinetic temperature profile, which is approximated by a two-step power law of the form $T_\mathrm{g,1}(r) = T_\star\ r^{-\epsilon_1}$ for $r \leq R_\mathrm{t}$ and $T_\mathrm{g,2}(r) = T_\mathrm{g,1}(R_\mathrm{t})\ r^{-\epsilon_2}$ for $r \geq R_\mathrm{t}$. We vary $\epsilon_1$ and $\epsilon_2$ from 0.0 to 1.1 and the transition radius $R_\mathrm{t}$ from 5 R$_\star$ to 50 R$_\star$. A power law with $\epsilon = 0.5$ for the gas kinetic temperature is expected for optically thin regions \citep{dec2006}, but we allow for significantly steeper laws as well in view of the high optical depth in \oh's CSE.

We use the spectrally resolved low-J CO transitions observed with JCMT and HIFI to constrain the free parameters. Following \citet{dec2007}, the evaluation of the model grid is done in two steps. First, all models that do not agree with the absolute flux calibration uncertainties $\sigma_\mathrm{abs}$ on the data sets, as specified in Sect.~\ref{sect:data}, are excluded. Then, a goodness-of-fit assessment based on the log-likelihood function is set up to judge the shape of the line profile, taking statistical noise $\sigma_\mathrm{stat}$ into account. For this last step, a scaling factor is introduced to equalize the integrated intensity of the observed line profile with the integrated intensity of the predicted line profile. The JCMT data do not significantly detect the CO $J=6-5$ transition. We use both the $3\sigma_\mathrm{stat}$ noise level and $\sigma_\mathrm{abs}$ to define an upper limit for the predicted intensities of this line. We also compare the predicted line profiles of the CO $J=2-1$ and $J=3-2$ JCMT observations with the soft parabola component of the fitted line profile, rather than the observed line profile, in which the interstellar CO contamination does not allow for a reliable determination of the integrated intensity and the line profile shape.

\begin{figure}[!t]
\begin{center}$
\begin{array}{c}
\resizebox{8.7cm}{!}{\includegraphics{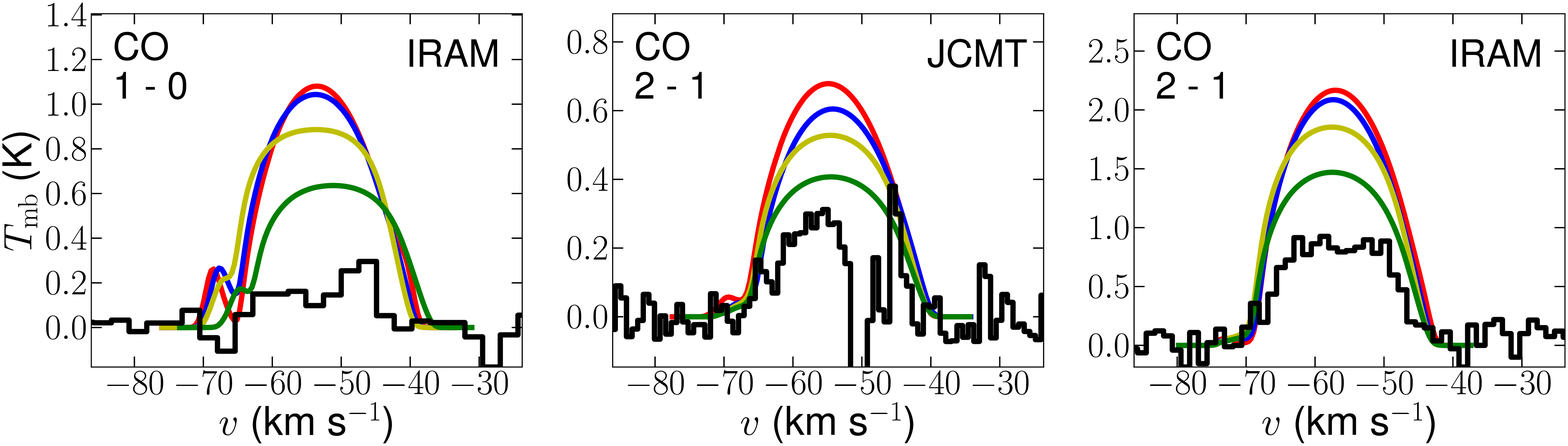}} \\
\resizebox{5.8cm}{!}{\includegraphics{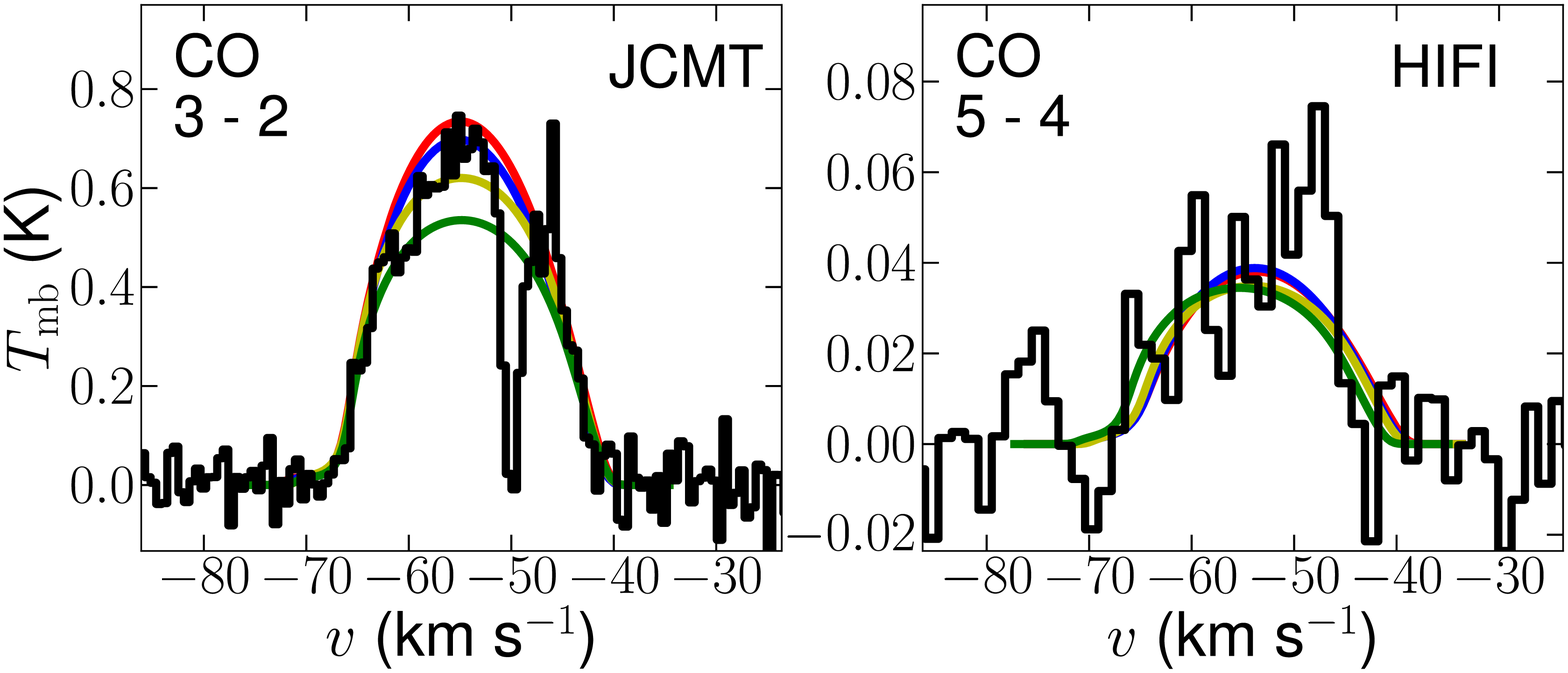}} \\
\resizebox{5.8cm}{!}{\includegraphics{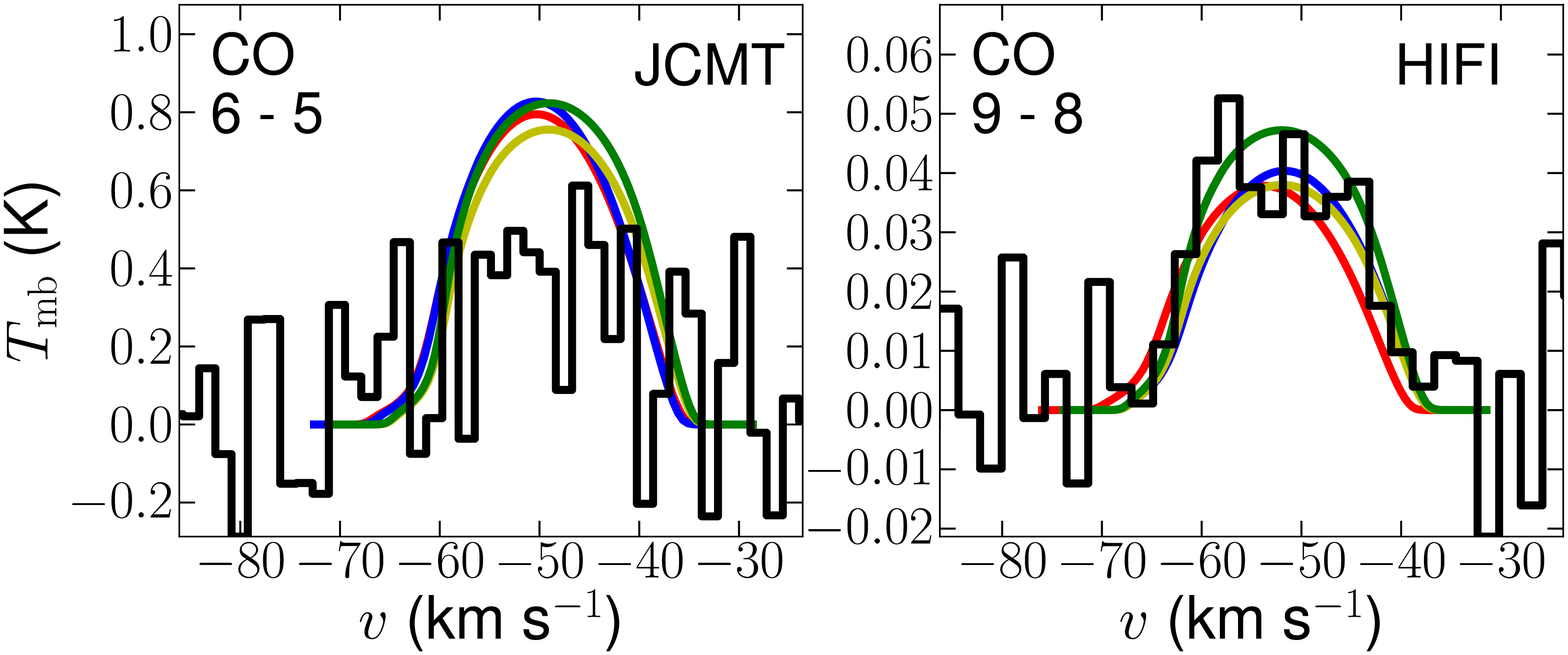}} \\
\end{array}$
\end{center}
\vspace{-0.5cm}
\caption{The spectrally resolved low-J CO observations of \oh are shown in black. The colored curves correspond to the models listed in Table~\ref{table:templaws}, which assume a constant mass-loss rate: 1.~red; 2.~blue; 3.~yellow; 4.~green. See Sect.~\ref{sect:valco} for further discussion of the validity of these CO models.}
\label{fig:models}
\end{figure}
With the exception of the CO $J=1-0$ and $J=2-1$ observations, four models reproduce all of the available CO transitions, shown in Fig.~\ref{fig:models}. Our estimate of the uncertainty on the mass-loss rates given in Table~\ref{table:templaws} amounts to a factor of three on the given values and is dominated {by the sampling resolution of the mass-loss-rate parameter in the model grid, as well as by the uncertainty of the CO abundance that we assume.} These values compare well with the mass-loss-rate estimates of $\mg \sim 5\times10^{-5}\ \msunyr$ reported in the two most recent studies that included \oh \citep{suh2002b,deb2010}. Because the CO lines in the PACS wavelength region are undetected, the PACS data only provide an upper limit for the high-J CO emission lines. All models listed in Table~\ref{table:templaws} agree with this upper limit.
\begin{table}[!t]
	{
    	\setlength{\tabcolsep}{5pt}
    	\caption{Values for the grid parameters of the four best fit models to the CO molecular emission data. Listed are the stellar effective temperature $T_\star$, the powers of the 2-step power law $\epsilon_1$ and $\epsilon_2$, the transition radius $R_\mathrm{t}$, the gas-mass-loss rate $\mg$, the dust-to-gas-ratio \psiemp, and the critical \water abundance \waterabuncrit.}\label{table:templaws}
  	\begin{center}
   	\begin{tabular}[c]{clllllll}\hline\hline\rule[0mm]{0mm}{3mm}
 &  $T_\star$   &  $\epsilon_1$     & $\epsilon_2$     & $R_\mathrm{t}$ & $\mg$ & {\psiemp} & {\waterabuncrit} \\ 
 &(K)&&&  (R$_\star$) & ($\msunyr$)  & & \\ \hline
1  & 3500	& 0.2	& 0.9 	& 5 & $1.0 \times 10^{-4}$ & {0.005}& {$8.5 \times 10^{-5}$} \\
2  & 3000	& 0.2	& 0.9	& 5 & $5.0 \times 10^{-5}$ & {0.01} & {$1.7 \times 10^{-4}$} \\
3  & 2500	& 0.2	& 0.9	& 5 & $2.0 \times 10^{-5}$ & {0.025}& {$4.0 \times 10^{-4}$} \\
4  & 2000	& 0.01	& 1.0	& 5 & $2.0 \times 10^{-5}$ & {0.025}& {$4.0 \times 10^{-4}$} \\
	\hline
   	\end{tabular}
   	\end{center}
	}
\end{table}
\subsubsection{Validity of CO model results}\label{sect:valco}
Model 1 in Table~\ref{table:templaws} requires a stellar effective temperature of 3500 K, which is comparatively high for OH/IR stars. Owing to the high optical thickness of the circumstellar shells in OH/IR stars, the common method of deriving stellar effective temperatures based on V-K color measurements cannot be used to constrain the effective temperature \citep{deb2010}. \citet{lep1995} have attempted to constrain the effective temperatures for a large sample of OH/IR stars based on near-IR (K-L') colors. They find temperatures lower than 3000 K for the whole sample, contrasting with the value found for our Model 1. We choose not to exclude Model 1 because of the uncertainty involved in determining effective temperatures for sources with optically thick shells.

All predictions in Table~\ref{table:templaws} overestimate the CO $J=2-1$ observations by a factor 1.5 up to 3 and the CO $J=1-0$ line by a factor of 3 up to 5. Two explanations are possible:
\begin{enumerate}
 \item The CO $J=1-0$ and $J=2-1$ lines are formed in the outermost part of the CSE, where the contribution of the interstellar radiation field cannot be neglected. This radiation field depends strongly on the local conditions. For instance, if a strong UV-source is present near \oh, the photodissociation radius of CO determined from the general formalism derived by \citet{mam1988} would decrease. Reducing $\rog \sim 50 \times 10^3\ \mathrm{R}_\star$ to $\rog \sim 1500$ - $2000\ R_\star$ would allow the model to predict the observed intensity of the CO $J=1-0$ and $J=2-1$ lines correctly, while keeping the intensity of the higher-J lines the same. However, this is remarkably close to the radius of the OH 1612 MHz maser shell in \oh, which \citet{bow1990} found to be $(1.38 \pm 0.14)''$. This translates to $\mathrm{r}_\mathrm{OH} \sim 1000$ - $2000\ R_\star$ at a distance of 2.1 kpc, depending on the assumed temperature at the stellar surface. This suggests that such a small outer CSE radius is unlikely for \oh.
 \item The mass loss in \oh may be variable, as suggested by several previous studies (e.g.~\citeauthor{deb2010}~\citeyear{deb2010}). If the mass-loss rate has been lower in the past, then the low-J lines might have a lower intensity compared to our predictions assuming a constant mass-loss rate. To improve the prediction of the $J=1-0$ and $J=2-1$ CO lines, we calculated models with a change in mass-loss rate going from $\mg = 1 \times 10^{-7}\ \msunyr$ in the outer wind up to $\mg$ as listed in Table~\ref{table:templaws} for the inner wind. The transition from high to low mass-loss rate occurs gradually at the radial distance $R_\mathrm{VM}$ of $\sim 2500-4000\ R_\star$, which translates to $\sim 7.5-14.5 \times 10^{16}$ cm. \citet{del1997} found similar results based on the IRAM $^{12}$CO and $^{13}$CO $J=2-1$ and $J=1-0$ transitions with an older, low mass-loss rate of $\mgl  < 5 \times 10^{-6}\ \msunyr$ and a recent, high $\mgh$ between $5 \times 10^{-5}$ and $5 \times 10^{-4}\ \msunyr$. They found a transitional radius of $R_\mathrm{VM} \sim 1.8 - 5.3\ \times 10^{16}$ cm, depending on $\mgh$. Our estimate of $R_\mathrm{VM}$ is larger, but we have a stronger constraint on $R_\mathrm{VM}$ due to the higher-J CO transitions. The values we find for $R_\mathrm{VM}$ translate to an increase in the mass-loss rate in \oh in the last 2000 up to 4000 years, depending on $\mgh$ and the temperature structure. This recent change in mass-loss rate is commonly referred to as the recent onset of the superwind, which is often suggested for many OH/IR stars by several studies (\citeauthor{jus1992}~\citeyear{jus1992}, \citeauthor{del1997}~\citeyear{del1997}, de Vries et al. in prep, Justtanont et al. in prep).
\end{enumerate}
The assumption of a change in mass-loss rate to predict the low-J CO line strengths correctly does not affect further modeling of other emission lines, as long as these lines originate in a region within the radial distance $R_\mathrm{VM}$. This is the case for the \water vapor emission lines detected in the PACS wavelength range, so we use the four models listed in Table~\ref{table:templaws} in what follows.

\subsubsection{\water emission}\label{sect:wateremis}
To determine the \water vapor abundance, we use \psiemp and adopt the gas kinetic temperature law and gas-mass-loss rate of Model 2 in Table~\ref{table:templaws} because the mass-loss rate is closest to the estimates of previous studies. What follows has been done for every model in Table~\ref{table:templaws}, and even though the resulting values scale with the mass-loss rate, the general conclusions do not change.

We have selected 18 mostly unblended, non-masing \water emission lines in the PACS spectrum to fit the \gastronoom models. The selection of lines is indicated in Table~\ref{table:intint}. We assume an ortho-to-para \water ratio ($OPR$) of 3 \citep{dec2010b}. When using \psiemp $ =0.01$ derived from fitting CO emission and the thermal IR continuum (see Sect.~\ref{sect:d2gemp}) for Model 2 in Table~\ref{table:templaws}, we find an unexpectedly low \water vapor abundance\footnote{\water vapor abundances are always given for ortho-\water alone, while \water column densities and \water ice abundances always include both ortho- and para-\water.} \waterabun $\sim 5 \times 10^{-6}$, as compared with \waterabun $\sim 3 \times 10^{-4}$ derived from chemical models \citep{che2006}. \citet{mae2008} also found an \water vapor abundance of $\sim 10^{-6}$ for the OH/IR source WX Psc, indicating that such a discrepancy has been found before in sources that have a high mass-loss rate. 

\begin{figure}[!t]
\resizebox{\hsize}{!}{\includegraphics{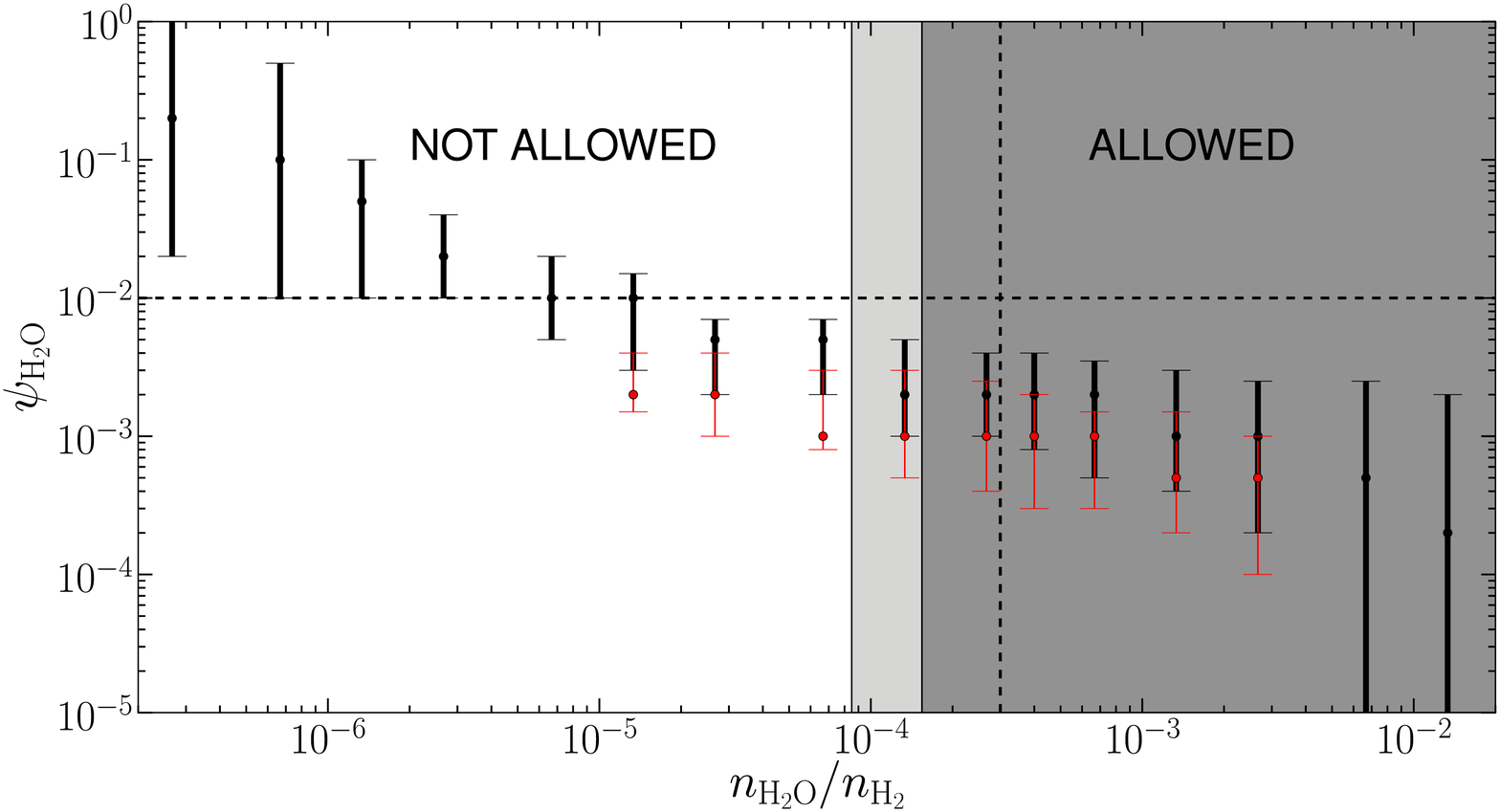}}
\caption{\oh \water emission spectrum modeling results for the temperature law and mass-loss rate of Models 1 and 2 in Table~\ref{table:templaws} in red and black, respectively. \psiwater and its uncertainty is determined for a wide range of (ortho + para) \water vapor abundances. From the modeling of the IR continuum and the CO data, a value of \psiemp$= 0.01$ is determined. The expected \water vapor abundance from chemical models is $3 \times 10^{-4}$ \citep{che2006}. Both values are indicated by the dashed black lines. The dark gray area indicates the lower limit defined by the critical \water vapor abundance derived from the \water ice fraction of Model 2, see Sect.~\ref{sect:vaporice}. For comparison, the light gray area indicates the lower limit found for Model 1.}
\label{fig:d2g_vs_h2o}
\end{figure}

To resolve this discrepancy, we determine \psiwater for a wide range of \water vapor abundances such that our model reproduces the \water emission spectrum of \oh. The results for Model 2 in Table~\ref{table:templaws} are shown in Fig.~\ref{fig:d2g_vs_h2o} and give further clues to the excitation mechanism of \water vapor in the high mass-loss-rate case. At values $\geq 10^{-3}$, \psiwater correlates with the \water vapor abundance. Here, pumping through excitation by the dust radiation field plays an important role. For lower dust-to-gas ratios, the dust radiation field becomes negligible for \water vapor excitation causing the correlation between \psiwater and \waterabun to level off. The correlation between \psiwater and the \water vapor abundance depends on the gas-mass-loss rate. For comparison, equivalent results for Model 1 in Table~\ref{table:templaws} are shown in Fig.~\ref{fig:d2g_vs_h2o}.

Figures \ref{fig:pacs1} and \ref{fig:pacs2} show the continuum-subtracted PACS spectrum compared to the predictions of Model 2 in Table~\ref{table:templaws} for \waterabun $=3\times10^{-4}$ and \psiwater $=0.003$. Included and indicated on the spectrum are all $^{12}$CO rotational transitions in the vibrational groundstate and all o-\water and p-\water transitions in the vibrational groundstate and the $\nu_1=1$ and $\nu_2=1$ vibrational states with rotational quantum number up to $J_\mathrm{upper}=8$ in the PACS wavelength range, regardless of being detected or not. The 18 \water transitions used in the initial fitting procedure are indicated as well. We calculated model spectra for the other temperature and density profiles in Table~\ref{table:templaws} and arrive at the same overall result as for Model 2 with some small differences in the relative line strengths of the lines.

\subsubsection{Validity of \water model results}
\begin{figure*}
$
\begin{array}{cc}
\resizebox{8.8cm}{!}{\includegraphics{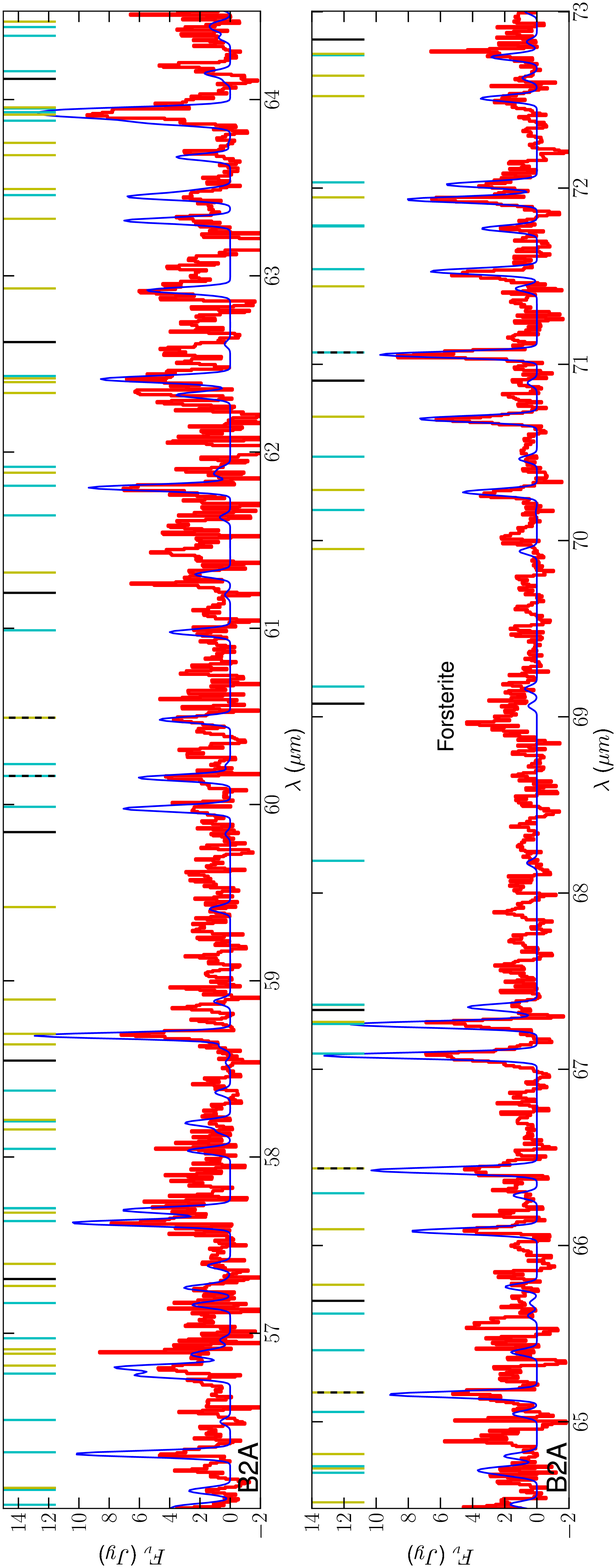}} & \resizebox{8.8cm}{!}{\includegraphics{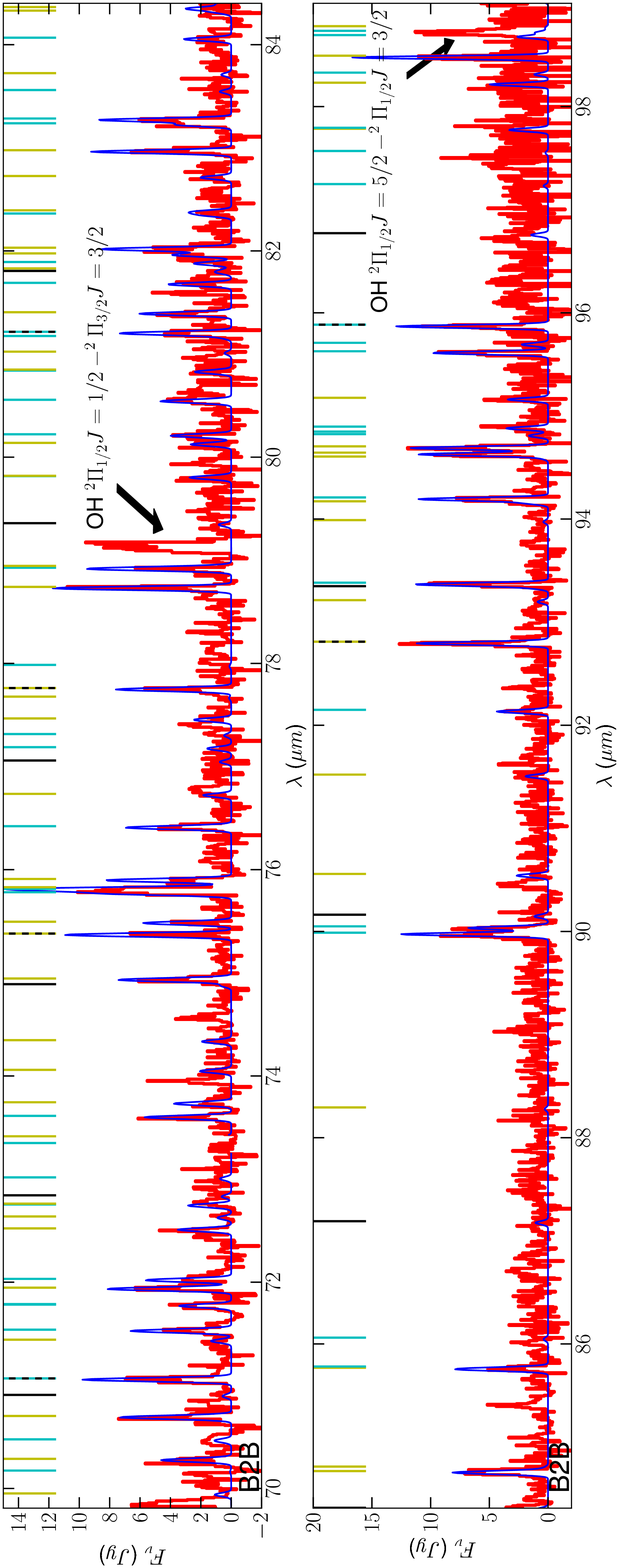}} \\
\end{array}$
\caption{The continuum-subtracted PACS spectrum of \oh is shown in red for the blue bands. The PACS band is indicated in the lower left corner of each spectrum. Model 2 in Table~\ref{table:templaws} with \waterabun$= 3 \times 10^{-4}$ and \psiwater $=0.003$ is given in blue. The other parameters are listed in Tables~\ref{table:inputpar} and \ref{table:modelpar}. The colored vertical lines indicate the molecule contributing at that specific wavelength, with full black for \co, yellow for ortho-\water, and cyan for para-\water. The dashed black-colored lines indicate the water lines used for the initial \water line fitting. The forsterite feature at $\sim 69$ \mic (not completely removed during continuum subtraction) and the OH rotational cascade lines $^2\Pi_{1/2} J=1/2 - ^2\Pi_{3/2} J=3/2$ and $^2\Pi_{1/2} J=5/2 - ^2\Pi_{1/2} J=3/2$ at $\sim 79.1$ \mic and $\sim 98.7$ \mic, respectively, (not included in our modeling) are labeled.}
\label{fig:pacs1}
\end{figure*}

\begin{figure*}
$
\begin{array}{cc}
\resizebox{8.5cm}{!}{\includegraphics{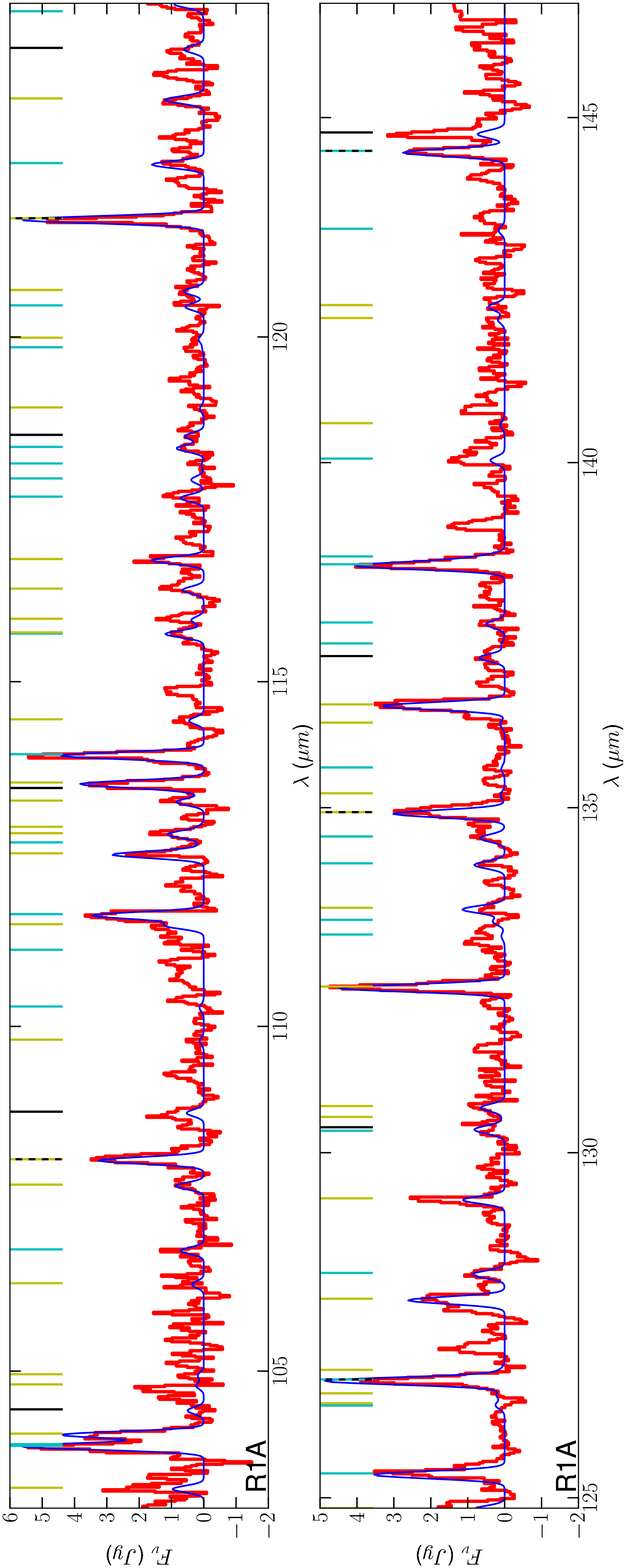}} & \resizebox{8.5cm}{!}{\includegraphics{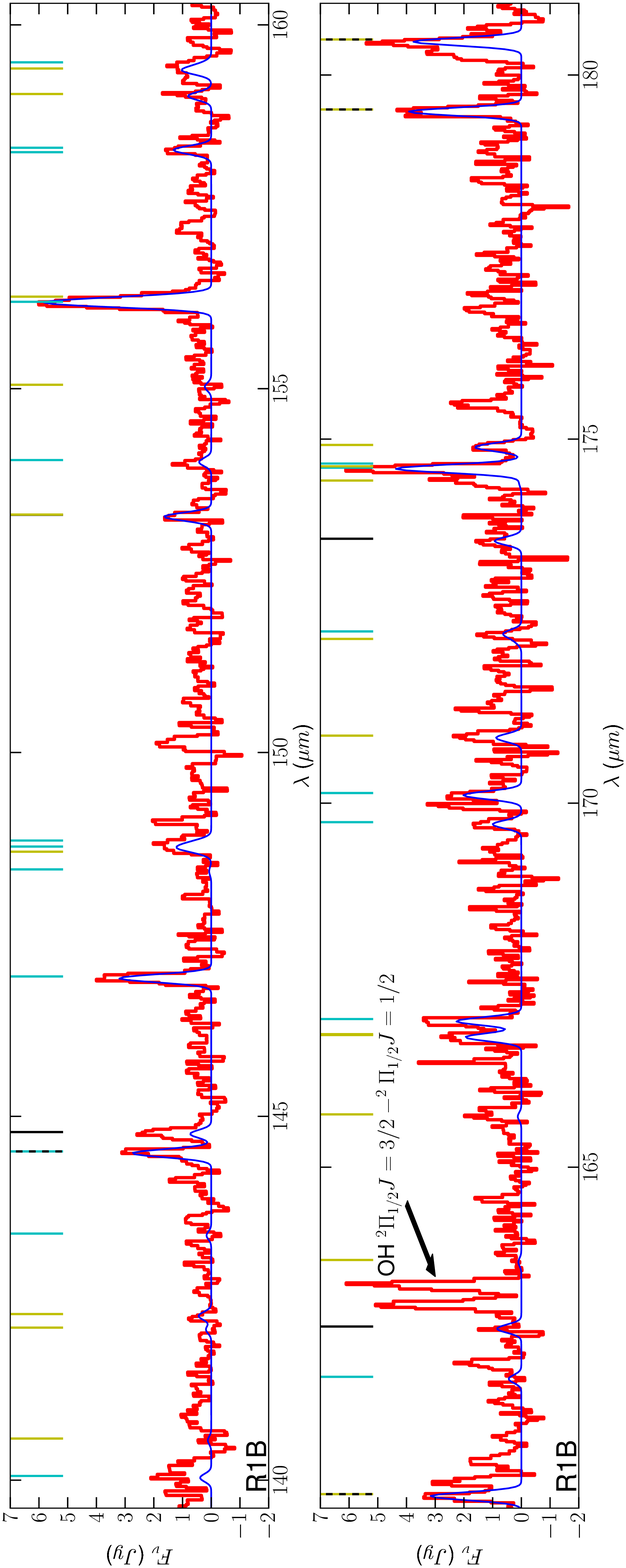}} \\
\end{array}$
\caption{The continuum-subtracted PACS spectrum of \oh is shown in red for the red bands. The PACS band is indicated in the lower left corner of each spectrum. Model 2 in Table~\ref{table:templaws} with \waterabun$= 3 \times 10^{-4}$ and \psiwater $=0.003$ is given in blue. The other parameters are listed in Tables~\ref{table:inputpar} and \ref{table:modelpar}. The colored vertical lines indicate the molecule contributing at that specific wavelength, with full black for \co, yellow for ortho-\water, and cyan for para-\water. The dashed black-colored lines indicate the water lines used for the initial \water line fitting. The OH rotational cascade line $^2\Pi_{1/2} J=3/2 - ^2\Pi_{1/2} J=1/2$ at $\sim 162.9$ \mic (not included in our modeling) is labeled. At $\sim 144.9$ \mic, another strong line appears both in band R1A and band R1B. A different wavelength sampling causes the line in band R1B to appear weaker, but the integrated line fluxes of both lines are within the absolute flux calibration uncertainty of PACS. This line remains unidentified. The CO line alone cannot explain the observed integrated line flux at this wavelength.}
\label{fig:pacs2}
\end{figure*}

A slight downward trend is present in the comparison between model predictions and the observations, as shown in Figs.~\ref{fig:pacs1} and \ref{fig:pacs2}, with a systematic overestimation at short wavelengths and a systematic underestimation at longer wavelengths. This difference is within the 30\% absolute uncertainty calibration of the PACS data. However, a relative trend between short and long wavelengths in the model-to-data comparison is unexpected from the absolute calibration errors. A relative uncertainty between short and long wavelengths can be caused by pointing errors of the telescope, but this effect is likely too small to explain the trend that we find. This trend is present for all models in Table~\ref{table:templaws}, although less evident for Models 3 and 4 (with the lower mass-loss-rate estimate of $\mg \sim 2 \times 10^{-5}\ \msunyr$).

Based on this, one could opt to exclude Models 1 and 2. However, \water is not a good tracer of the density and temperature structure owing to the complexity of \water excitation mechanisms and possible maser effects. Normally, CO is a good density and temperature tracer, because CO is dominated by collisional excitation and does not mase. However, for \oh, CO lines are optically thick and were not reliably detected in the PACS wavelength range. In this case, $^{13}$CO lines would be a better tracer, but they are significantly weaker than $^{12}$CO emission lines and, as such, are not detected at all in the PACS observations. As long as the majority of \water lines are reproduced well over a wide range of wavelengths in the PACS data, for which the signal-to-noise ratio is low especially at short wavelengths, we consider a model to be satisfying. Thus, we choose not to exclude any models based on the trend in the predictions.

This large a set of \water lines has not been modeled before in such detail, covering full radiative transfer modeling of the CSE of a high mass-loss-rate OH/IR star. The consistent prediction of line-integrated fluxes of \water lines across a wide wavelength range that is well within the absolute flux calibration of the PACS instrument - especially in the red bands - is remarkable, considering the large number of \water lines and the complexity of the problem.

\subsubsection{The \water vapor-ice connection}\label{sect:vaporice}
\begin{figure}[!t]
\resizebox{\hsize}{!}{\includegraphics{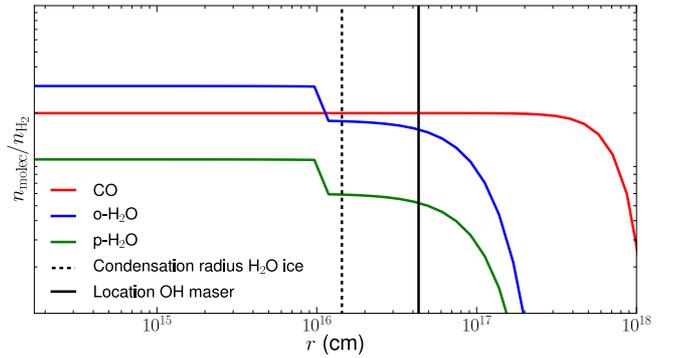}}
\caption{Schematic representation of the CO (red), the ortho-\water (blue), and para-\water (green) abundance profiles. The vertical dashed black line indicates the \water ice condensation radius. The vertical full black line indicates the location of the OH 1612 MHz maser shell, assuming a distance of 2100 pc. The signal-to-noise of the PACS is too low to trace the drop in \water vapor abundance (shown here for a freeze-out of $\sim 40\%$) at the \water ice condensation radius.}
\label{fig:water_abundance}
\end{figure}

An additional constraint {can be placed on the estimate of} the \water vapor abundance and the associated \psiwater. The presence of \water ice in an OH/IR CSE provides a lower limit on the \water vapor abundance. The condensation temperature of \water ice is $T_\mathrm{cond,ice} = 110$ K, following \citet{kam2009}. The condensation radius associated with 110 K is $R_\mathrm{cond,ice} = 1.2 \times 10^{16}$ cm. The line formation region for all unblended, nonmasing \water vapor lines in the spectrum of \oh is mostly within this radius. We can therefore define a critical \water vapor abundance at $r < R_{\mathrm{cond,ice}}$, below which there would not be enough \water vapor to form the observed amount of \water ice at $R_{\mathrm{cond,ice}}$. Following our modeling of the \water ice feature, the \water ice column density at $r>R_\mathrm{cond,ice}$ is $8.3\times 10^{17}$ cm$^{-2}$, which leads to a critical (ortho + para) \water vapor abundance of \waterabuncrit $= (1.7 \pm 0.2) \times 10^{-4}$. This critical abundance depends on the gas-mass-loss rate, as the ice mass is compared to the equivalent molecular hydrogen mass in the ice shell. As such, this critical value will be different for the three mass-loss rates given in Table~\ref{table:templaws}. For comparison, the critical \water vapor abundance is shown in Fig.~\ref{fig:d2g_vs_h2o} for Models 1 and 2 in Table~\ref{table:templaws}. The work by \citet{dij2003} suggests that at most 20\% of the \water vapor will freeze out onto dust grains. The actual \water vapor abundance {is thus expected to be} larger than the critical \water vapor abundance. Figure \ref{fig:water_abundance} gives a schematic representation of what the \water vapor abundance profile might look like, taking \water ice condensation into account (with a freeze-out of $\sim 40 \%$, a value that is arbitrarily chosen) and \water vapor photodissociation in the outer envelope.

Observational evidence of a larger actual \water vapor abundance than the critical \water vapor abundance is given by the presence of emission from the OH maser at 1612 MHz in a shell at $\mathrm{r} > \mathrm{R}_{\mathrm{cond,ice}}$. The photodissociation of \water into OH and H is one of the main OH production paths, throughout the whole envelope, as long as interstellar UV radiation is available to break up \water molecules. \citet{net1987} have shown that the OH abundance reaches a maximum at the radial distance where the OH maser shell at 1612 MHz is observed, indicating that other methods of OH production closer to the star, such as shock chemistry, can be ignored. As a result, \water needs to be present in the CSE at least up to the radial distance where the OH abundance peaks. In the case of \oh, the radius of the OH 1612 MHz maser shell is $(1.38 \pm 0.14)''$ \citep{bow1990}, which translates to $\mathrm{r}_\mathrm{OH} = (4.3 \pm 0.4) \times 10^{16}$ cm at a distance of 2.1 kpc. \citet{net1987} also give a formula for the expected OH 1612 MHz maser shell radius, which depends on the assumed interstellar radiation field \citep{hab1968}. Assuming the average Habing field, we find $6.1 \times 10^{16}$ cm, whereas the high Habing field leads to a shell radius of $4.3 \times 10^{16}$ cm. 


Our results for the critical \water vapor abundance agree well with those found in other studies. \citet{che2006} derived the expected abundances for several molecules from thermodynamic equilibrium and shock-induced NLTE chemistry, and found \waterabun $ \sim 3.0 \times 10^{-4}$ in O-rich AGB stars. \water vapor abundances derived by \citet{mae2008} for most sources in their sample lie between \waterabun $= 2.0 \times 10^{-4}$ and $1.5 \times 10^{-3}$. They do find a remarkably low \water vapor abundance of $\sim 10^{-6}$ for the OH/IR star WX Psc and offer two explanations: 1) \water ice formation depletes \water in gaseous form, and 2) \water lines may be formed in a region of a more recent, lower mass-loss rate. However, \citet{mae2008} use \psiemp for their molecular emission modeling. Our modeling has indicated that \psiemp leads to too low an \water abundance in \oh when compared to the \water ice content. Given that \oh and WX Psc are similar, their low value for the \water vapor abundance in WX Psc could also be the result of the use of \psiemp{ as an estimate of the dust-to-gas ratio}.
\subsection{Discussion: The dust-to-gas ratio}\label{sect:discussion}
The dust-to-gas ratio typical of AGB circumstellar environments is 0.005 \citep{whi1994}. We derive different values depending on the method used (for Model 2 in Table~\ref{table:templaws}): 
\begin{enumerate}
 \item We find \psiemp $= 0.01$, {accurate to within {a factor of three}}, from IR continuum and CO molecular emission modeling.
 \item The momentum transfer equation leads {to a dust-to-gas ratio \psimom $= 0.0005$,} while assuming complete momentum coupling between dust and gas, i.e.~dust grains of every grain size are coupled to the gas. For circumstellar environments typical of stars like \oh, \citet{mac1992} find that silicate grains with initial size smaller than $\sim 0.05$ \mic decouple from the gas near the condensation radius. If the coupling is not complete, a higher dust content is required to arrive at the same kinematical structure of the envelope, implying that \psimom $\geq 0.0005$.
 \item The critical \water vapor abundance provides a strong constraint on the expected initial \water vapor abundance in \oh. Applying our value of \waterabuncrit$> 1.7 \times 10^{-4}$ to the grid calculation shown in Fig.~\ref{fig:d2g_vs_h2o}, we find an upper limit for the associated dust-to-gas ratio \psiwater$ < 0.005$. This upper limit takes the uncertainty shown in Fig.~\ref{fig:d2g_vs_h2o} into account and assumes the unlikely case of 100\% freeze-out of \water vapor into \water ice. For reference, assuming a freeze-out of 20\%, we arrive at \psiwaterfo$\sim 0.0015$, accurate to within a factor of two.
\end{enumerate}

\begin{figure}[!t]
\resizebox{\hsize}{!}{\includegraphics{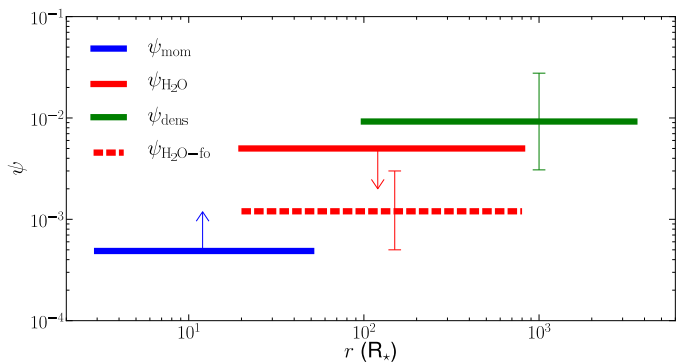}}
\caption{The results of determining the dust-to-gas ratio using the three different methods described in Sect.~\ref{sect:d2gemp} are shown {for Model 2 in Table~\ref{table:templaws}}. The horizontal bar indicates the part of the envelope traced by the method. The vertical bar indicates the uncertainty on the indicated value. \psimom is a lower limit, whereas \psiwater is an upper limit. {For reference, the dashed red line indicates \psiwaterfo assuming 20\% freeze-out of \water vapor into \water ice.} {The relative differences between the three values of the dust-to-gas ratio for the other models in Table~\ref{table:templaws} are similar, but scale upward or downward uniformly depending on the gas-mass-loss rate.} See Sect.~\ref{sect:discussion} for a more detailed description.}
\label{fig:d2gfinal}
\end{figure}

The results obtained for the dust-to-gas ratio appear incompatible. However, each method traces a different part of the envelope (see Fig.~\ref{fig:d2gfinal}).
\begin{enumerate}
 \item \psiemp is based on modeling the thermal dust emission, which traces the dust content of the envelope out to a radius of $\sim 5000\ \mathrm{R}_\star$, and the CO $J = 9-8$ down to $J=3-2$ emission lines. These lines are formed in the outer regions of the CSE at $100$ R$_\star < r < 4000$ R$_\star$, for Model 2 in Table~\ref{table:templaws}. Assuming the dust-mass-loss-rate remains constant throughout the whole CSE, \psiemp is therefore sensitive only to the outermost region of the envelope.
 \item \psimom is determined from the momentum transfer equation and therefore traces the acceleration zone, which in our model is located at $r<50$ R$_\star$.
 \item \psiwater traces the outflow at $20$ R$_\star <r<800$ R$_\star$ where all of the \water emission lines used to determine the \water abundance are formed.
\end{enumerate}
As shown in Fig.~\ref{fig:d2gfinal}, our findings tentatively point to the presence of a gradient in the dust-to-gas ratio with radial distance. {The results shown here are for Model 2 in Table~\ref{table:templaws}, but the same relative differences between the dust-to-gas ratio estimates are seen for the other models in Table~\ref{table:templaws}.} There could be several potential explanations for such behavior.

First, we assume a constant mass-loss rate for both the gas and dust components. If \oh's mass-loss history is not constant, a recent increase in the gas-mass-loss rate can explain the gradient in the dust-to-gas ratio only when the dust-mass-loss rate has not increased by the same factor as well. This is possible only if the dust forms less efficiently for an increased gas density. There is no immediate evidence that suggests such behavior for higher mass-loss rates, so this scenario appears to be unlikely.

Second, 84\% of the dust mass is formed in the innermost region of the envelope, at a few stellar radii, and the dust-mass-loss rate is assumed to be constant. If dust formation extends beyond the vicinity of the dust condensation radius, this could explain the gradient in the dust-to-gas ratio. \water ice formation is possible at a radius of $\sim 1000\ R_\star$ in the case of Model 2 in Table~\ref{table:templaws}, owing to the high \water vapor abundance. However, the amount of \water ice formed is not enough to explain the radial increase in dust-to-gas ratio. Formation of other dust species (such as silicates) at large distances from the star is unlikely due to the lower densities of the precursor molecular species when compared to \water vapor.

Third, we do not take clumping into account in the models. If clumps are present in the envelope, the ones close to the stellar surface are likely to be much more optically thick than those in the outer envelope. As a result, we trace the real amount of gas and dust in the outer envelope, whereas we trace a seemingly lower amount of gas and dust in the inner envelope. If clumps are responsible for the gradient in the dust-to-gas ratio, we have to assume that the optical depth effect caused by clumping is more severe for dust than for gas. Considering that a cloud of gas particles experiences an internal pressure, whereas a cloud of dust particles does not, this could be a valid assumption. We note that a clumped wind is also invoked by \citet{dij2006} to explain the observed high crystalline \water ice fraction in \oh.

\section{Conclusions}\label{sect:conclusion}
We have combined two state-of-the-art radiative transfer codes, \mcmax for the continuum radiative transfer, and \gastronoom for the line radiative transfer. We justified the use of more consistent dust properties in the gas modeling by showing that the dust component of the CSE has a significant influence on the excitation of \water at high mass-loss rates, while the dust condensation radius is important for both CO and \water at low mass-loss rates. 

We presented new PACS data of \oh, the first AGB OH/IR star for which a far-IR spectrum was taken with this instrument. We applied our approach to the combination of the PACS spectrum, HIFI observations of two CO transitions taken in the framework of the SUCCESS Herschel Guaranteed Time Program, ground-based JCMT observations of low-J CO transitions, and the ISO-SWS and ISO-LWS spectra. The combination of the HIFI and ground-based observations suggests a discrepancy between the lowest-J ($J=1-0$ and $J=2-1$) and the higher-J ($J=3-2$ and up) CO lines, which may point to a recent onset of a superwind in \oh. The IR continuum is modeled with a dust composition of metallic iron, amorphous silicates, crystalline silicates (forsterite and enstatite), and amorphous and crystalline \water ice. We found a dust-mass-loss rate of $\md=(5 \pm 1)\times 10^{-7}\ \msunyr$ and a contribution of \water ice to the total amount of dust beyond the \water ice condensation radius of $(16 \pm 2)\%$ with a crystalline-to-amorphous ratio of $0.8\pm0.2$. The CO transitions are modeled with an empirical temperature law resulting in four models with a constant gas-mass-loss rate ranging between $\mg =  1.0 \times 10^{-4}\ \msunyr$ and $\mg =  0.2 \times 10^{-4}\ \msunyr$, accurate to within a factor of three. The older mass-loss episode, traced by the outer regions of the CSE, is estimated to be $\mg \sim 1 \times 10^{-7}\ \msunyr$ with the transition between the low and high mass-loss rate occurring at $R_\mathrm{VM} \sim 2500-4000\ R_\star$. We derived a critical \water vapor abundance of $(1.7\pm0.2) \times 10^{-4}$ from the \water ice content of the CSE. This constrains the minimum amount of \water vapor required to produce the observed amount of \water ice assuming $100\%$ freeze-out efficiency. We note that the comparison between \water vapor models and the PACS spectrum shows a flux overestimation at shorter wavelengths and a flux underestimation at longer wavelengths. Even though these differences are within the absolute flux calibration of the PACS instrument, the wavelength-dependent discrepancy cannot be explained.

We derived the dust-to-gas ratio following three methods, which are sensitive to different regions of the outflow. We found for the first time indications of a gradient in the dust-to-gas ratio with radial distance from the star. Possible explanations for this behavior can include clumpiness, variable mass loss, or continued dust growth beyond the condensation radius, of which the first suggestion seems the most likely. 

Additionally, we reported the first detection in an AGB circumstellar environment of OH cascade rotational lines involved in the far-infrared pumping mechanism of the 1612 MHz OH maser. 

\begin{acknowledgements}
We would like to thank B. Acke and A.J.~van Marle for their contribution to the study. We also express gratitude toward the referee, who provided instructive feedback. RL acknowledges financial support from the Fund for Scientific Research - Flanders (FWO) under grant number ZKB5757-04-W01, and from the Department of Physics and Astronomy of the KULeuven. LD, EDB, and BdV acknowledge financial support from the FWO. JB and PR acknowledge support from the Belgian Federal Science Policy Office via the PRODEX Program of ESA. PACS has been developed by a consortium of institutes led by MPE (Germany) and including UVIE (Austria); KUL, CSL, IMEC (Belgium); CEA, OAMP (France); MPIA (Germany); IFSI, OAP/AOT, OAA/CAISMI, LENS, SISSA (Italy); IAC (Spain). This development has been supported by the funding agencies BMVIT (Austria), ESA-PRODEX (Belgium), CEA/CNES (France), DLR (Germany), ASI (Italy), and CICT/MCT (Spain). For the computations we used the infrastructure of the VSC - Flemish Supercomputer Center, funded by the Hercules Foundation and the Flemish Government - department EWI.
\end{acknowledgements}

\bibliographystyle{aa}
\bibliography{allreferences}

\begin{thebibliography}{81}
\expandafter\ifx\csname natexlab\endcsname\relax\def\natexlab#1{#1}\fi

\bibitem[{{Arenou} {et~al.}(1992){Arenou}, {Grenon}, \& {Gomez}}]{are1992}
{Arenou}, F., {Grenon}, M., \& {Gomez}, A. 1992, \aap, 258, 104

\bibitem[{{Barlow} {et~al.}(1996){Barlow}, {Nguyen-Q-Rieu}, {Truong-Bach},
  {Cernicharo}, {Gonzalez-Alfonso}, {Liu}, {Cox}, {Sylvester}, {Clegg},
  {Griffin}, {Swinyard}, {Unger}, {Baluteau}, {Caux}, {Cohen}, {Cohen},
  {Emery}, {Fischer}, {Furniss}, {Glencross}, {Greenhouse}, {Gry}, {Joubert},
  {Lim}, {Lorenzetti}, {Nisini}, {Omont}, {Orfei}, {Pequignot}, {Saraceno},
  {Serra}, {Skinner}, {Smith}, {Walker}, {Armand}, {Burgdorf}, {Ewart}, {di
  Giorgio}, {Molinari}, {Price}, {Sidher}, {Texier}, \& {Trams}}]{bar1996}
{Barlow}, M.~J., {Nguyen-Q-Rieu}, {Truong-Bach}, {et~al.} 1996, \aap, 315, L241

\bibitem[{{Beck} {et~al.}(1992){Beck}, {Gail}, {Henkel}, \&
  {Sedlmayr}}]{bec1992}
{Beck}, H.~K.~B., {Gail}, H.-P., {Henkel}, R., \& {Sedlmayr}, E. 1992, \aap,
  265, 626

\bibitem[{{Bertie} {et~al.}(1969){Bertie}, {Labb{\'e}}, \& {Whalley}}]{ber1969}
{Bertie}, J.~E., {Labb{\'e}}, H.~J., \& {Whalley}, E. 1969, \jcp, 50, 4501

\bibitem[{{Bohren} \& {Huffman}(1983)}]{boh1983}
{Bohren}, C.~F. \& {Huffman}, D.~R. 1983, {Absorption and scattering of light
  by small particles}, ed. C.~F. {Bohren} \& D.~R. {Huffman}

\bibitem[{{Bowers} \& {Johnston}(1990)}]{bow1990}
{Bowers}, P.~F. \& {Johnston}, K.~J. 1990, \apj, 354, 676

\bibitem[{{Bujarrabal} \& {Alcolea}(1991)}]{buj1991}
{Bujarrabal}, V. \& {Alcolea}, J. 1991, \aap, 251, 536

\bibitem[{{Cherchneff}(2006)}]{che2006}
{Cherchneff}, I. 2006, \aap, 456, 1001

\bibitem[{{Chiar} \& {Tielens}(2006)}]{chi2006}
{Chiar}, J.~E. \& {Tielens}, A.~G.~G.~M. 2006, \apj, 637, 774

\bibitem[{{Clegg} {et~al.}(1996){Clegg}, {Ade}, {Armand}, {Baluteau}, {Barlow},
  {Buckley}, {Berges}, {Burgdorf}, {Caux}, {Ceccarelli}, {Cerulli}, {Church},
  {Cotin}, {Cox}, {Cruvellier}, {Culhane}, {Davis}, {di Giorgio}, {Diplock},
  {Drummond}, {Emery}, {Ewart}, {Fischer}, {Furniss}, {Glencross},
  {Greenhouse}, {Griffin}, {Gry}, {Harwood}, {Hazell}, {Joubert}, {King},
  {Lim}, {Liseau}, {Long}, {Lorenzetti}, {Molinari}, {Murray}, {Naylor},
  {Nisini}, {Norman}, {Omont}, {Orfei}, {Patrick}, {Pequignot}, {Pouliquen},
  {Price}, {Nguyen-Q-Rieu}, {Rogers}, {Robinson}, {Saisse}, {Saraceno},
  {Serra}, {Sidher}, {Smith}, {Smith}, {Spinoglio}, {Swinyard}, {Texier},
  {Towlson}, {Trams}, {Unger}, \& {White}}]{cle1996}
{Clegg}, P.~E., {Ade}, P.~A.~R., {Armand}, C., {et~al.} 1996, \aap, 315, L38

\bibitem[{{De Beck} {et~al.}(2010){De Beck}, {Decin}, {de Koter}, {Justtanont},
  {Verhoelst}, {Kemper}, \& {Menten}}]{deb2010}
{De Beck}, E., {Decin}, L., {de Koter}, A., {et~al.} 2010, \aap, 523, A18

\bibitem[{{de Graauw} {et~al.}(1996){de Graauw}, {Haser}, {Beintema},
  {Roelfsema}, {van Agthoven}, {Barl}, {Bauer}, {Bekenkamp}, {Boonstra},
  {Boxhoorn}, {Cote}, {de Groene}, {van Dijkhuizen}, {Drapatz}, {Evers},
  {Feuchtgruber}, {Frericks}, {Genzel}, {Haerendel}, {Heras}, {van der Hucht},
  {van der Hulst}, {Huygen}, {Jacobs}, {Jakob}, {Kamperman}, {Katterloher},
  {Kester}, {Kunze}, {Kussendrager}, {Lahuis}, {Lamers}, {Leech}, {van der
  Lei}, {van der Linden}, {Luinge}, {Lutz}, {Melzner}, {Morris}, {van Nguyen},
  {Ploeger}, {Price}, {Salama}, {Schaeidt}, {Sijm}, {Smoorenburg}, {Spakman},
  {Spoon}, {Steinmayer}, {Stoecker}, {Valentijn}, {Vandenbussche}, {Visser},
  {Waelkens}, {Waters}, {Wensink}, {Wesselius}, {Wiezorrek}, {Wieprecht},
  {Wijnbergen}, {Wildeman}, \& {Young}}]{deg1996}
{de Graauw}, T., {Haser}, L.~N., {Beintema}, D.~A., {et~al.} 1996, \aap, 315,
  L49

\bibitem[{{de Graauw} {et~al.}(2010){de Graauw}, {Helmich}, {Phillips},
  {Stutzki}, {Caux}, {Whyborn}, {Dieleman}, {Roelfsema}, {Aarts}, {Assendorp},
  {Bachiller}, {Baechtold}, {Barcia}, {Beintema}, {Belitsky}, {Benz}, {Bieber},
  {Boogert}, {Borys}, {Bumble}, {Ca{\"i}s}, {Caris}, {Cerulli-Irelli},
  {Chattopadhyay}, {Cherednichenko}, {Ciechanowicz}, {Coeur-Joly}, {Comito},
  {Cros}, {de Jonge}, {de Lange}, {Delforges}, {Delorme}, {den Boggende},
  {Desbat}, {Diez-Gonz{\'a}lez}, {di Giorgio}, {Dubbeldam}, {Edwards},
  {Eggens}, {Erickson}, {Evers}, {Fich}, {Finn}, {Franke}, {Gaier}, {Gal},
  {Gao}, {Gallego}, {Gauffre}, {Gill}, {Glenz}, {Golstein}, {Goulooze},
  {Gunsing}, {G{\"u}sten}, {Hartogh}, {Hatch}, {Higgins}, {Honingh}, {Huisman},
  {Jackson}, {Jacobs}, {Jacobs}, {Jarchow}, {Javadi}, {Jellema}, {Justen},
  {Karpov}, {Kasemann}, {Kawamura}, {Keizer}, {Kester}, {Klapwijk}, {Klein},
  {Kollberg}, {Kooi}, {Kooiman}, {Kopf}, {Krause}, {Krieg}, {Kramer},
  {Kruizenga}, {Kuhn}, {Laauwen}, {Lai}, {Larsson}, {Leduc}, {Leinz}, {Lin},
  {Liseau}, {Liu}, {Loose}, {L{\'o}pez-Fernandez}, {Lord}, {Luinge}, {Marston},
  {Mart{\'{\i}}n-Pintado}, {Maestrini}, {Maiwald}, {McCoey}, {Mehdi}, {Megej},
  {Melchior}, {Meinsma}, {Merkel}, {Michalska}, {Monstein}, {Moratschke},
  {Morris}, {Muller}, {Murphy}, {Naber}, {Natale}, {Nowosielski}, {Nuzzolo},
  {Olberg}, {Olbrich}, {Orfei}, {Orleanski}, {Ossenkopf}, {Peacock}, {Pearson},
  {Peron}, {Phillip-May}, {Piazzo}, {Planesas}, {Rataj}, {Ravera}, {Risacher},
  {Salez}, {Samoska}, {Saraceno}, {Schieder}, {Schlecht}, {Schl{\"o}der},
  {Schm{\"u}lling}, {Schultz}, {Schuster}, {Siebertz}, {Smit}, {Szczerba},
  {Shipman}, {Steinmetz}, {Stern}, {Stokroos}, {Teipen}, {Teyssier}, {Tils},
  {Trappe}, {van Baaren}, {van Leeuwen}, {van de Stadt}, {Visser}, {Wildeman},
  {Wafelbakker}, {Ward}, {Wesselius}, {Wild}, {Wulff}, {Wunsch}, {Tielens},
  {Zaal}, {Zirath}, {Zmuidzinas}, \& {Zwart}}]{deg2010}
{de Graauw}, T., {Helmich}, F.~P., {Phillips}, T.~G., {et~al.} 2010, \aap, 518,
  L6

\bibitem[{{de Vries} {et~al.}(2010){de Vries}, {Min}, {Waters}, {Blommaert}, \&
  {Kemper}}]{dev2010}
{de Vries}, B.~L., {Min}, M., {Waters}, L.~B.~F.~M., {Blommaert}, J.~A.~D.~L.,
  \& {Kemper}, F. 2010, \aap, 516, A86

\bibitem[{{Decin} {et~al.}(2010{\natexlab{a}}){Decin}, {De Beck},
  {Br{\"u}nken}, {M{\"u}ller}, {Menten}, {Kim}, {Willacy}, {de Koter}, \&
  {Wyrowski}}]{dec2010a}
{Decin}, L., {De Beck}, E., {Br{\"u}nken}, S., {et~al.} 2010{\natexlab{a}},
  \aap, 516, A69

\bibitem[{{Decin} \& {Eriksson}(2007)}]{dec2007b}
{Decin}, L. \& {Eriksson}, K. 2007, \aap, 472, 1041

\bibitem[{{Decin} {et~al.}(2006){Decin}, {Hony}, {de Koter}, {Justtanont},
  {Tielens}, \& {Waters}}]{dec2006}
{Decin}, L., {Hony}, S., {de Koter}, A., {et~al.} 2006, \aap, 456, 549

\bibitem[{{Decin} {et~al.}(2007){Decin}, {Hony}, {de Koter}, {Molenberghs},
  {Dehaes}, \& {Markwick-Kemper}}]{dec2007}
{Decin}, L., {Hony}, S., {de Koter}, A., {et~al.} 2007, \aap, 475, 233

\bibitem[{{Decin} {et~al.}(2010{\natexlab{b}}){Decin}, {Justtanont}, {De Beck},
  {Lombaert}, {de Koter}, {Waters}, {Marston}, {Teyssier}, {Sch{\"o}ier},
  {Bujarrabal}, {Alcolea}, {Cernicharo}, {Dominik}, {Melnick}, {Menten},
  {Neufeld}, {Olofsson}, {Planesas}, {Schmidt}, {Szczerba}, {de Graauw},
  {Helmich}, {Roelfsema}, {Dieleman}, {Morris}, {Gallego},
  {D{\'{\i}}ez-Gonz{\'a}lez}, \& {Caux}}]{dec2010b}
{Decin}, L., {Justtanont}, K., {De Beck}, E., {et~al.} 2010{\natexlab{b}},
  \aap, 521, L4

\bibitem[{{Delfosse} {et~al.}(1997){Delfosse}, {Kahane}, \&
  {Forveille}}]{del1997}
{Delfosse}, X., {Kahane}, C., \& {Forveille}, T. 1997, \aap, 320, 249

\bibitem[{{Dijkstra} {et~al.}(2006){Dijkstra}, {Dominik}, {Bouwman}, \& {de
  Koter}}]{dij2006}
{Dijkstra}, C., {Dominik}, C., {Bouwman}, J., \& {de Koter}, A. 2006, \aap,
  449, 1101

\bibitem[{{Dijkstra} {et~al.}(2003){Dijkstra}, {Dominik}, {Hoogzaad}, {de
  Koter}, \& {Min}}]{dij2003}
{Dijkstra}, C., {Dominik}, C., {Hoogzaad}, S.~N., {de Koter}, A., \& {Min}, M.
  2003, \aap, 401, 599

\bibitem[{{Elitzur} {et~al.}(1976){Elitzur}, {Goldreich}, \&
  {Scoville}}]{eli1976}
{Elitzur}, M., {Goldreich}, P., \& {Scoville}, N. 1976, \apj, 205, 384

\bibitem[{{Engels} {et~al.}(1986){Engels}, {Schmid-Burgk}, \&
  {Walmsley}}]{eng1986}
{Engels}, D., {Schmid-Burgk}, J., \& {Walmsley}, C.~M. 1986, \aap, 167, 129

\bibitem[{{Gilman}(1969)}]{gil1969}
{Gilman}, R.~C. 1969, \apjl, 155, L185

\bibitem[{{Gray} {et~al.}(2005){Gray}, {Howe}, \& {Lewis}}]{gra2005}
{Gray}, M.~D., {Howe}, D.~A., \& {Lewis}, B.~M. 2005, \mnras, 364, 783

\bibitem[{{Groenewegen} {et~al.}(2011){Groenewegen}, {Waelkens}, {Barlow},
  {Kerschbaum}, {Garcia-Lario}, {Cernicharo}, {Blommaert}, {Bouwman}, {Cohen},
  {Cox}, {Decin}, {Exter}, {Gear}, {Gomez}, {Hargrave}, {Henning},
  {Hutsem{\'e}kers}, {Ivison}, {Jorissen}, {Krause}, {Ladjal}, {Leeks}, {Lim},
  {Matsuura}, {Naz{\'e}}, {Olofsson}, {Ottensamer}, {Polehampton}, {Posch},
  {Rauw}, {Royer}, {Sibthorpe}, {Swinyard}, {Ueta}, {Vamvatira-Nakou},
  {Vandenbussche}, {van de Steene}, {van Eck}, {van Hoof}, {van Winckel},
  {Verdugo}, \& {Wesson}}]{gro2011}
{Groenewegen}, M.~A.~T., {Waelkens}, C., {Barlow}, M.~J., {et~al.} 2011, \aap,
  526, A162

\bibitem[{{Gustafsson} {et~al.}(2008){Gustafsson}, {Edvardsson}, {Eriksson},
  {J{\o}rgensen}, {Nordlund}, \& {Plez}}]{gus2008}
{Gustafsson}, B., {Edvardsson}, B., {Eriksson}, K., {et~al.} 2008, \aap, 486,
  951

\bibitem[{{Habing}(1968)}]{hab1968}
{Habing}, H.~J. 1968, \bain, 19, 421

\bibitem[{{Habing} \& {Olofsson}(2003)}]{hab2003}
{Habing}, H.~J. \& {Olofsson}, H., eds. 2003, {Asymptotic giant branch stars}

\bibitem[{{Henning} \& {Stognienko}(1996)}]{hen1996}
{Henning}, T. \& {Stognienko}, R. 1996, \aap, 311, 291

\bibitem[{{Herman} \& {Habing}(1985)}]{her1985}
{Herman}, J. \& {Habing}, H.~J. 1985, \aaps, 59, 523

\bibitem[{{Heske} {et~al.}(1990){Heske}, {Forveille}, {Omont}, {van der Veen},
  \& {Habing}}]{hes1990}
{Heske}, A., {Forveille}, T., {Omont}, A., {van der Veen}, W.~E.~C.~J., \&
  {Habing}, H.~J. 1990, \aap, 239, 173

\bibitem[{{H{\"o}fner}(2008)}]{hof2008}
{H{\"o}fner}, S. 2008, \aap, 491, L1

\bibitem[{{J{\"a}ger} {et~al.}(1998){J{\"a}ger}, {Molster}, {Dorschner},
  {Henning}, {Mutschke}, \& {Waters}}]{jag1998a}
{J{\"a}ger}, C., {Molster}, F.~J., {Dorschner}, J., {et~al.} 1998, \aap, 339,
  904

\bibitem[{{Justtanont} \& {Tielens}(1992)}]{jus1992}
{Justtanont}, K. \& {Tielens}, A.~G.~G.~M. 1992, \apj, 389, 400

\bibitem[{{Kama} {et~al.}(2009){Kama}, {Min}, \& {Dominik}}]{kam2009}
{Kama}, M., {Min}, M., \& {Dominik}, C. 2009, \aap, 506, 1199

\bibitem[{{Kemper} {et~al.}(2002){Kemper}, {de Koter}, {Waters}, {Bouwman}, \&
  {Tielens}}]{kem2002}
{Kemper}, F., {de Koter}, A., {Waters}, L.~B.~F.~M., {Bouwman}, J., \&
  {Tielens}, A.~G.~G.~M. 2002, \aap, 384, 585

\bibitem[{{Kemper} {et~al.}(2003){Kemper}, {Stark}, {Justtanont}, {de Koter},
  {Tielens}, {Waters}, {Cami}, \& {Dijkstra}}]{kem2003}
{Kemper}, F., {Stark}, R., {Justtanont}, K., {et~al.} 2003, \aap, 407, 609

\bibitem[{{Kessler} {et~al.}(1996){Kessler}, {Steinz}, {Anderegg}, {Clavel},
  {Drechsel}, {Estaria}, {Faelker}, {Riedinger}, {Robson}, {Taylor}, \&
  {Xim{\'e}nez de Ferr{\'a}n}}]{kes1996}
{Kessler}, M.~F., {Steinz}, J.~A., {Anderegg}, M.~E., {et~al.} 1996, \aap, 315,
  L27

\bibitem[{{Knapp} \& {Morris}(1985)}]{kna1985}
{Knapp}, G.~R. \& {Morris}, M. 1985, \apj, 292, 640

\bibitem[{{Kwok}(1975)}]{kwo1975}
{Kwok}, S. 1975, \apj, 198, 583

\bibitem[{{Lamers} \& {Cassinelli}(1999)}]{lam1999}
{Lamers}, H.~J.~G.~L.~M. \& {Cassinelli}, J.~P. 1999, {Introduction to Stellar
  Winds} (Cambridge University Press)

\bibitem[{{Lepine} {et~al.}(1995){Lepine}, {Ortiz}, \& {Epchtein}}]{lep1995}
{Lepine}, J.~R.~D., {Ortiz}, R., \& {Epchtein}, N. 1995, \aap, 299, 453

\bibitem[{{Loup} {et~al.}(1993){Loup}, {Forveille}, {Omont}, \&
  {Paul}}]{lou1993}
{Loup}, C., {Forveille}, T., {Omont}, A., \& {Paul}, J.~F. 1993, \aaps, 99, 291

\bibitem[{{MacGregor} \& {Stencel}(1992)}]{mac1992}
{MacGregor}, K.~B. \& {Stencel}, R.~E. 1992, \apj, 397, 644

\bibitem[{{Maercker} {et~al.}(2009){Maercker}, {Sch{\"o}ier}, {Olofsson},
  {Bergman}, {Frisk}, {.~Hjalmarson}, {Justtanont}, {Kwok}, {Larsson},
  {Olberg}, \& {Sandqvist}}]{mae2009a}
{Maercker}, M., {Sch{\"o}ier}, F.~L., {Olofsson}, H., {et~al.} 2009, \aap, 494,
  243

\bibitem[{{Maercker} {et~al.}(2008){Maercker}, {Sch{\"o}ier}, {Olofsson},
  {Bergman}, \& {Ramstedt}}]{mae2008}
{Maercker}, M., {Sch{\"o}ier}, F.~L., {Olofsson}, H., {Bergman}, P., \&
  {Ramstedt}, S. 2008, \aap, 479, 779

\bibitem[{{Mamon} {et~al.}(1988){Mamon}, {Glassgold}, \& {Huggins}}]{mam1988}
{Mamon}, G.~A., {Glassgold}, A.~E., \& {Huggins}, P.~J. 1988, \apj, 328, 797

\bibitem[{{Mathis} {et~al.}(1977){Mathis}, {Rumpl}, \& {Nordsieck}}]{mat1977}
{Mathis}, J.~S., {Rumpl}, W., \& {Nordsieck}, K.~H. 1977, \apj, 217, 425

\bibitem[{{Menten} \& {Melnick}(1989)}]{men1989}
{Menten}, K.~M. \& {Melnick}, G.~J. 1989, \apjl, 341, L91

\bibitem[{{Menten} {et~al.}(2006){Menten}, {Philipp}, {G{\"u}sten}, {Alcolea},
  {Polehampton}, \& {Br{\"u}nken}}]{men2006}
{Menten}, K.~M., {Philipp}, S.~D., {G{\"u}sten}, R., {et~al.} 2006, \aap, 454,
  L107

\bibitem[{{Min} {et~al.}(2009){Min}, {Dullemond}, {Dominik}, {de Koter}, \&
  {Hovenier}}]{min2009a}
{Min}, M., {Dullemond}, C.~P., {Dominik}, C., {de Koter}, A., \& {Hovenier},
  J.~W. 2009, \aap, 497, 155

\bibitem[{{Min} {et~al.}(2003){Min}, {Hovenier}, \& {de Koter}}]{min2003}
{Min}, M., {Hovenier}, J.~W., \& {de Koter}, A. 2003, \aap, 404, 35

\bibitem[{{Netzer} \& {Knapp}(1987)}]{net1987}
{Netzer}, N. \& {Knapp}, G.~R. 1987, \apj, 323, 734

\bibitem[{{Neufeld} {et~al.}(1996){Neufeld}, {Chen}, {Melnick}, {de Graauw},
  {Feuchtgruber}, {Haser}, {Lutz}, \& {Harwit}}]{neu1996}
{Neufeld}, D.~A., {Chen}, W., {Melnick}, G.~J., {et~al.} 1996, \aap, 315, L237

\bibitem[{{Norris} {et~al.}(2012){Norris}, {Tuthill}, {Ireland}, {Lacour},
  {Zijlstra}, {Lykou}, {Evans}, {Stewart}, \& {Bedding}}]{nor2012}
{Norris}, B.~R.~M., {Tuthill}, P.~G., {Ireland}, M.~J., {et~al.} 2012, \nat,
  484, 220

\bibitem[{{Olofsson}(2008)}]{olo2008}
{Olofsson}, H. 2008, Physica Scripta Volume T, 133, 014028

\bibitem[{{Omont} {et~al.}(1990){Omont}, {Forveille}, {Moseley}, {Glaccum},
  {Harvey}, {Likkel}, {Loewenstein}, \& {Lisse}}]{omo1990}
{Omont}, A., {Forveille}, T., {Moseley}, S.~H., {et~al.} 1990, \apjl, 355, L27

\bibitem[{{Pilbratt} {et~al.}(2010){Pilbratt}, {Riedinger}, {Passvogel},
  {Crone}, {Doyle}, {Gageur}, {Heras}, {Jewell}, {Metcalfe}, {Ott}, \&
  {Schmidt}}]{pil2010}
{Pilbratt}, G.~L., {Riedinger}, J.~R., {Passvogel}, T., {et~al.} 2010, \aap,
  518, L1

\bibitem[{{Poglitsch} {et~al.}(2010){Poglitsch}, {Waelkens}, {Geis},
  {Feuchtgruber}, {Vandenbussche}, {Rodriguez}, {Krause}, {Renotte}, {van
  Hoof}, {Saraceno}, {Cepa}, {Kerschbaum}, {Agn{\`e}se}, {Ali}, {Altieri},
  {Andreani}, {Augueres}, {Balog}, {Barl}, {Bauer}, {Belbachir}, {Benedettini},
  {Billot}, {Boulade}, {Bischof}, {Blommaert}, {Callut}, {Cara}, {Cerulli},
  {Cesarsky}, {Contursi}, {Creten}, {De Meester}, {Doublier}, {Doumayrou},
  {Duband}, {Exter}, {Genzel}, {Gillis}, {Gr{\"o}zinger}, {Henning},
  {Herreros}, {Huygen}, {Inguscio}, {Jakob}, {Jamar}, {Jean}, {de Jong},
  {Katterloher}, {Kiss}, {Klaas}, {Lemke}, {Lutz}, {Madden}, {Marquet},
  {Martignac}, {Mazy}, {Merken}, {Montfort}, {Morbidelli}, {M{\"u}ller},
  {Nielbock}, {Okumura}, {Orfei}, {Ottensamer}, {Pezzuto}, {Popesso},
  {Putzeys}, {Regibo}, {Reveret}, {Royer}, {Sauvage}, {Schreiber}, {Stegmaier},
  {Schmitt}, {Schubert}, {Sturm}, {Thiel}, {Tofani}, {Vavrek}, {Wetzstein},
  {Wieprecht}, \& {Wiezorrek}}]{pog2010}
{Poglitsch}, A., {Waelkens}, C., {Geis}, N., {et~al.} 2010, \aap, 518, L2

\bibitem[{{Redfield} \& {Linsky}(2008)}]{red2008}
{Redfield}, S. \& {Linsky}, J.~L. 2008, \apj, 673, 283

\bibitem[{{Renzini}(1981)}]{ren1981}
{Renzini}, A. 1981, in Astrophysics and Space Science Library, Vol.~88,
  Physical Processes in Red Giants, ed. {I.~Iben Jr.~\& A.~Renzini}, 431--446

\bibitem[{{Russell}(1934)}]{rus1934}
{Russell}, H.~N. 1934, \apj, 79, 317

\bibitem[{{Servoin} \& {Piriou}(1973)}]{ser1973}
{Servoin}, J.~L. \& {Piriou}, B. 1973, Phys. Status Solidi (B), 55, 677

\bibitem[{{Skinner} {et~al.}(1999){Skinner}, {Justtanont}, {Tielens}, {Betz},
  {Boreiko}, \& {Baas}}]{ski1999}
{Skinner}, C.~J., {Justtanont}, K., {Tielens}, A.~G.~G.~M., {et~al.} 1999,
  \mnras, 302, 293

\bibitem[{{Sloan} {et~al.}(2003){Sloan}, {Kraemer}, {Price}, \&
  {Shipman}}]{slo2003}
{Sloan}, G.~C., {Kraemer}, K.~E., {Price}, S.~D., \& {Shipman}, R.~F. 2003,
  \apjs, 147, 379

\bibitem[{{Suh}(2004)}]{suh2004}
{Suh}, K. 2004, \apj, 615, 485

\bibitem[{{Suh} \& {Kim}(2002)}]{suh2002b}
{Suh}, K. \& {Kim}, H. 2002, \aap, 391, 665

\bibitem[{{Swinyard} {et~al.}(1996){Swinyard}, {Clegg}, {Ade}, {Armand},
  {Baluteau}, {Barlow}, {Berges}, {Burgdorf}, {Caux}, {Ceccarelli}, {Cerulli},
  {Church}, {Colgan}, {Cotin}, {Cox}, {Cruvellier}, {Davis}, {Digiorgio},
  {Emery}, {Ewart}, {Fischer}, {Furniss}, {Glencross}, {Greenhouse}, {Griffin},
  {Gry}, {Haas}, {Joubert}, {King}, {Lim}, {Liseau}, {Lord}, {Lorenzetti},
  {Molinari}, {Naylor}, {Nisini}, {Omont}, {Orfei}, {Patrick}, {Pequignot},
  {Pouliquen}, {Price}, {Nguyen-Q-Rieu}, {Robinson}, {Saisse}, {Saraceno},
  {Serra}, {Sidher}, {Smith}, {Spinoglio}, {Texier}, {Towlson}, {Trams},
  {Unger}, \& {White}}]{swi1996}
{Swinyard}, B.~M., {Clegg}, P.~E., {Ade}, P.~A.~R., {et~al.} 1996, \aap, 315,
  L43

\bibitem[{{Sylvester} {et~al.}(1997){Sylvester}, {Barlow}, {Nguyen-Q-Rieu},
  {Liu}, {Skinner}, {Cohen}, {Lim}, {Cox}, {Truong-Bach}, {Smith}, \&
  {Habing}}]{syl1997}
{Sylvester}, R.~J., {Barlow}, M.~J., {Nguyen-Q-Rieu}, {et~al.} 1997, \mnras,
  291, L42

\bibitem[{{Sylvester} {et~al.}(1999){Sylvester}, {Kemper}, {Barlow}, {de Jong},
  {Waters}, {Tielens}, \& {Omont}}]{syl1999}
{Sylvester}, R.~J., {Kemper}, F., {Barlow}, M.~J., {et~al.} 1999, \aap, 352,
  587

\bibitem[{{Tielens}(2005)}]{tie2005}
{Tielens}, A.~G.~G.~M. 2005, {The Physics and Chemistry of the Interstellar
  Medium} (Cambridge University Press)

\bibitem[{{Tielens} \& {Allamandola}(1987)}]{tie1987}
{Tielens}, A.~G.~G.~M. \& {Allamandola}, L.~J. 1987, in Astrophysics and Space
  Science Library, Vol. 134, Interstellar Processes, ed. {D.~J.~Hollenbach \&
  H.~A.~Thronson Jr.}, 397--469

\bibitem[{{Truong-Bach} {et~al.}(1991){Truong-Bach}, {Morris}, \&
  {Nguyen-Q-Rieu}}]{tru1991}
{Truong-Bach}, {Morris}, D., \& {Nguyen-Q-Rieu}. 1991, \aap, 249, 435

\bibitem[{{Truong-Bach} {et~al.}(1999){Truong-Bach}, {Sylvester}, {Barlow},
  {Nguyen-Q-Rieu}, {Lim}, {Liu}, {Baluteau}, {Deguchi}, {Justtanont}, \&
  {Tielens}}]{tru1999}
{Truong-Bach}, {Sylvester}, R.~J., {Barlow}, M.~J., {et~al.} 1999, \aap, 345,
  925

\bibitem[{{van Langevelde} {et~al.}(1990){van Langevelde}, {van der Heiden}, \&
  {van Schooneveld}}]{van1990}
{van Langevelde}, H.~J., {van der Heiden}, R., \& {van Schooneveld}, C. 1990,
  \aap, 239, 193

\bibitem[{{Warren}(1984)}]{war1984}
{Warren}, S.~G. 1984, \ao, 23, 1206

\bibitem[{{Whitelock} {et~al.}(1991){Whitelock}, {Feast}, \&
  {Catchpole}}]{whi1991}
{Whitelock}, P., {Feast}, M., \& {Catchpole}, R. 1991, \mnras, 248, 276

\bibitem[{{Whitelock} {et~al.}(1994){Whitelock}, {Menzies}, {Feast}, {Marang},
  {Carter}, {Roberts}, {Catchpole}, \& {Chapman}}]{whi1994}
{Whitelock}, P., {Menzies}, J., {Feast}, M., {et~al.} 1994, \mnras, 267, 711

\bibitem[{{Whittet}(1992)}]{whi1992}
{Whittet}, D.~C.~B. 1992, {Dust in the galactic environment}, ed. {Whittet,
  D.~C.~B.}

\end{thebibliography}

\appendix
\section{Integrated line strengths}
Table~\ref{table:intint} lists integrated line strengths of all detected ortho- and para-\water vapor emission lines and the 1612 MHz OH maser formation rotational cascade lines in the PACS spectrum shown in Figs.~\ref{fig:pacs1} and \ref{fig:pacs2}. Because the OH emission lines occur in doublets, the integrated line strengths for both components have been summed. We refer to \citet{syl1997} for details on OH spectroscopy. Where confusion due to line blending occurs, we indicate this clearly, as well as list all \water transitions that may contribute to the emission line. As such we cannot distinguish the relative contribution of each transition in the blend. Blends that might be caused by the emission of other molecules not modeled in this study are not indicated.
\onecolumn
\begin{longtab}
\begin{longtable}{llllrr}
\caption{{Integrated line strength $F_\mathrm{int}$ (W/m$^2$) of detected ortho- and para-\water vapor emission lines and the 1612 MHz OH maser formation rotational cascade lines in the PACS spectrum shown in Figs.~\ref{fig:pacs1} and \ref{fig:pacs2}. The rest wavelength $\lambda_0$ (\mic) of the transition is indicated. The percentages between brackets indicate the uncertainty on $F_\mathrm{int}$, which includes both the fitting uncertainty and the PACS absolute flux calibration uncertainty of 20\%.}}\label{table:intint}\\
\hline\hline
PACS&Molecule&Vibrational&Rotational&$\lambda_0$&\multicolumn{1}{c}{$F_\mathrm{int}$} \\
band&&state&transition&$\mu$m&\multicolumn{1}{c}{(W/m$^2$)} \\
\hline
\endfirsthead
\caption{continued.}\\
\hline\hline
PACS&Molecule&Vibrational&Rotational&$\lambda_0$&\multicolumn{1}{c}{$F_\mathrm{int}$} \\
band&&state&transition&$\mu$m&\multicolumn{1}{c}{(W/m$^2$)} \\
\hline
\endhead
\hline
\multicolumn{6}{c}{\parbox{\LTcapwidth}{\tablefoot{\tablefoottext{{$\star$}}{{Line strengths flagged for potential line blends (see Sect.~\ref{sect:pacs}). Transitions that might cause the line blend are mentioned immediately below the flagged transition.}}
\tablefoottext{{a}}{{Transition detected in an emission doublet. The given value is the sum of the line strengths of both emission lines in the doublet.}}
\tablefoottext{{b}}{{The selection of \water vapor emission lines based on which we have derived the \water vapor abundance (see Sect.~\ref{sect:wateremis}).}}
}}}
\endfoot
B2A&o-H$_2$O&$\nu=0$&$J_{\mathrm{K}_\mathrm{a}, \mathrm{K}_\mathrm{c}}=4_{3,2} - 3_{2,1}$&58.70&1.99e-16 (25.8\%)\\
&o-H$_2$O&$\nu_3=1$&$J_{\mathrm{K}_\mathrm{a}, \mathrm{K}_\mathrm{c}}=4_{3,1} - 3_{2,2}$&58.89&6.64e-17 (45.2\%)\\
&p-H$_2$O&$\nu=0$&$J_{\mathrm{K}_\mathrm{a}, \mathrm{K}_\mathrm{c}}=7_{2,6} - 6_{1,5}$&59.99&\tablefootmark{$\star$}1.51e-16 (38.6\%)\\
&p-H$_2$O&$\nu=0$&$J_{\mathrm{K}_\mathrm{a}, \mathrm{K}_\mathrm{c}}=8_{2,6} - 7_{3,5}$&60.16&\tablefootmark{b}8.90e-17 (34.8\%)\\
&o-H$_2$O&$\nu_2=1$&$J_{\mathrm{K}_\mathrm{a}, \mathrm{K}_\mathrm{c}}=3_{3,0} - 2_{2,1}$&60.49&\tablefootmark{b}8.65e-17 (40.6\%)\\
&o-H$_2$O&$\nu=0$&$J_{\mathrm{K}_\mathrm{a}, \mathrm{K}_\mathrm{c}}=6_{6,1} - 6_{5,2}$&63.91&\tablefootmark{$\star$}4.50e-16 (23.1\%)\\
&o-H$_2$O&$\nu=0$&$J_{\mathrm{K}_\mathrm{a}, \mathrm{K}_\mathrm{c}}=6_{6,0} - 6_{5,1}$&63.93&\\
&o-H$_2$O&$\nu_2=1$&$J_{\mathrm{K}_\mathrm{a}, \mathrm{K}_\mathrm{c}}=8_{0,8} - 7_{1,7}$&63.95&\\
&o-H$_2$O&$\nu=0$&$J_{\mathrm{K}_\mathrm{a}, \mathrm{K}_\mathrm{c}}=7_{6,1} - 7_{5,2}$&63.96&\\
&o-H$_2$O&$\nu=0$&$J_{\mathrm{K}_\mathrm{a}, \mathrm{K}_\mathrm{c}}=6_{2,5} - 5_{1,4}$&65.17&\tablefootmark{$\star$,b}1.31e-16 (35.2\%)\\
&o-H$_2$O&$\nu=0$&$J_{\mathrm{K}_\mathrm{a}, \mathrm{K}_\mathrm{c}}=3_{3,0} - 2_{2,1}$&66.44&\tablefootmark{b}1.28e-16 (29.0\%)\\
&p-H$_2$O&$\nu=0$&$J_{\mathrm{K}_\mathrm{a}, \mathrm{K}_\mathrm{c}}=3_{3,1} - 2_{2,0}$&67.09&\tablefootmark{$\star$}2.59e-16 (23.1\%)\\
&p-H$_2$O&$\nu_2=1$&$J_{\mathrm{K}_\mathrm{a}, \mathrm{K}_\mathrm{c}}=5_{2,4} - 4_{1,3}$&67.26&\tablefootmark{$\star$}1.52e-16 (25.2\%)\\
&p-H$_2$O&$\nu=0$&$J_{\mathrm{K}_\mathrm{a}, \mathrm{K}_\mathrm{c}}=3_{3,0} - 3_{0,3}$&67.27&\\
&o-H$_2$O&$\nu_2=1$&$J_{\mathrm{K}_\mathrm{a}, \mathrm{K}_\mathrm{c}}=3_{2,1} - 2_{1,2}$&70.29&7.59e-17 (27.8\%)\\
&o-H$_2$O&$\nu=0$&$J_{\mathrm{K}_\mathrm{a}, \mathrm{K}_\mathrm{c}}=8_{2,7} - 8_{1,8}$&70.70&1.31e-16 (22.4\%)\\
&p-H$_2$O&$\nu=0$&$J_{\mathrm{K}_\mathrm{a}, \mathrm{K}_\mathrm{c}}=5_{2,4} - 4_{1,3}$&71.07&\tablefootmark{b}1.92e-16 (22.6\%)\\
&p-H$_2$O&$\nu=0$&$J_{\mathrm{K}_\mathrm{a}, \mathrm{K}_\mathrm{c}}=7_{1,7} - 6_{0,6}$&71.54&1.11e-16 (29.2\%)\\
&o-H$_2$O&$\nu=0$&$J_{\mathrm{K}_\mathrm{a}, \mathrm{K}_\mathrm{c}}=7_{0,7} - 6_{1,6}$&71.95&1.28e-16 (26.3\%)\\\hline
B2B&o-H$_2$O&$\nu_2=1$&$J_{\mathrm{K}_\mathrm{a}, \mathrm{K}_\mathrm{c}}=3_{2,1} - 2_{1,2}$&70.29&\tablefootmark{$\star$}1.14e-16 (34.4\%)\\
&o-H$_2$O&$\nu=0$&$J_{\mathrm{K}_\mathrm{a}, \mathrm{K}_\mathrm{c}}=8_{2,7} - 8_{1,8}$&70.70&1.27e-16 (29.1\%)\\
&p-H$_2$O&$\nu=0$&$J_{\mathrm{K}_\mathrm{a}, \mathrm{K}_\mathrm{c}}=5_{2,4} - 4_{1,3}$&71.07&\tablefootmark{b}1.88e-16 (22.8\%)\\
&p-H$_2$O&$\nu=0$&$J_{\mathrm{K}_\mathrm{a}, \mathrm{K}_\mathrm{c}}=7_{1,7} - 6_{0,6}$&71.54&9.54e-17 (31.8\%)\\
&o-H$_2$O&$\nu=0$&$J_{\mathrm{K}_\mathrm{a}, \mathrm{K}_\mathrm{c}}=7_{0,7} - 6_{1,6}$&71.95&1.47e-16 (25.8\%)\\
&p-H$_2$O&$\nu=0$&$J_{\mathrm{K}_\mathrm{a}, \mathrm{K}_\mathrm{c}}=9_{3,7} - 9_{2,8}$&73.61&9.09e-17 (26.0\%)\\
&o-H$_2$O&$\nu=0$&$J_{\mathrm{K}_\mathrm{a}, \mathrm{K}_\mathrm{c}}=7_{2,5} - 6_{3,4}$&74.95&1.26e-16 (22.7\%)\\
&o-H$_2$O&$\nu=0$&$J_{\mathrm{K}_\mathrm{a}, \mathrm{K}_\mathrm{c}}=3_{2,1} - 2_{1,2}$&75.38&\tablefootmark{b}1.05e-16 (23.7\%)\\
&o-H$_2$O&$\nu=0$&$J_{\mathrm{K}_\mathrm{a}, \mathrm{K}_\mathrm{c}}=8_{5,4} - 8_{4,5}$&75.50&5.40e-17 (31.8\%)\\
&p-H$_2$O&$\nu=0$&$J_{\mathrm{K}_\mathrm{a}, \mathrm{K}_\mathrm{c}}=5_{5,1} - 5_{4,2}$&75.78&\tablefootmark{$\star$}3.11e-16 (21.1\%)\\
&p-H$_2$O&$\nu=0$&$J_{\mathrm{K}_\mathrm{a}, \mathrm{K}_\mathrm{c}}=7_{5,3} - 7_{4,4}$&75.81&\\
&p-H$_2$O&$\nu=0$&$J_{\mathrm{K}_\mathrm{a}, \mathrm{K}_\mathrm{c}}=6_{5,2} - 6_{4,3}$&75.83&\\
&o-H$_2$O&$\nu=0$&$J_{\mathrm{K}_\mathrm{a}, \mathrm{K}_\mathrm{c}}=5_{5,0} - 5_{4,1}$&75.91&\tablefootmark{$\star$}1.09e-16 (26.8\%)\\
&p-H$_2$O&$\nu=0$&$J_{\mathrm{K}_\mathrm{a}, \mathrm{K}_\mathrm{c}}=6_{5,1} - 6_{4,2}$&76.42&1.08e-16 (25.4\%)\\
&o-H$_2$O&$\nu=0$&$J_{\mathrm{K}_\mathrm{a}, \mathrm{K}_\mathrm{c}}=7_{5,2} - 7_{4,3}$&77.76&\tablefootmark{b}7.39e-17 (25.6\%)\\
&o-H$_2$O&$\nu=0$&$J_{\mathrm{K}_\mathrm{a}, \mathrm{K}_\mathrm{c}}=4_{2,3} - 3_{1,2}$&78.74&1.43e-16 (22.8\%)\\
&p-H$_2$O&$\nu=0$&$J_{\mathrm{K}_\mathrm{a}, \mathrm{K}_\mathrm{c}}=6_{1,5} - 5_{2,4}$&78.93&\tablefootmark{$\star$}8.61e-17 (26.6\%)\\
&p-H$_2$O&$\nu_2=1$&$J_{\mathrm{K}_\mathrm{a}, \mathrm{K}_\mathrm{c}}=6_{4,3} - 6_{3,4}$&78.95&\\
&OH&$\nu=0$&$^2\Pi_{1/2} J=1/2 - ^2\Pi_{3/2} J=3/2$&79.12-79.18&\tablefootmark{a}3.09e-16 (35.9\%) \\
&o-H$_2$O&$\nu=0$&$J_{\mathrm{K}_\mathrm{a}, \mathrm{K}_\mathrm{c}}=8_{3,6} - 8_{2,7}$&82.98&9.18e-17 (25.7\%)\\
&p-H$_2$O&$\nu_2=1$&$J_{\mathrm{K}_\mathrm{a}, \mathrm{K}_\mathrm{c}}=3_{2,2} - 2_{1,1}$&83.24&5.26e-17 (39.8\%)\\
&p-H$_2$O&$\nu=0$&$J_{\mathrm{K}_\mathrm{a}, \mathrm{K}_\mathrm{c}}=6_{0,6} - 5_{1,5}$&83.28&8.80e-17 (28.1\%)\\
&o-H$_2$O&$\nu=0$&$J_{\mathrm{K}_\mathrm{a}, \mathrm{K}_\mathrm{c}}=7_{1,6} - 7_{0,7}$&84.77&\tablefootmark{$\star$}1.56e-16 (22.9\%)\\
&o-H$_2$O&$\nu=0$&$J_{\mathrm{K}_\mathrm{a}, \mathrm{K}_\mathrm{c}}=8_{4,5} - 8_{3,6}$&85.77&\tablefootmark{$\star$}6.92e-17 (25.5\%)\\
&o-H$_2$O&$\nu_2=1$&$J_{\mathrm{K}_\mathrm{a}, \mathrm{K}_\mathrm{c}}=6_{4,2} - 6_{3,3}$&85.78&\\
&p-H$_2$O&$\nu=0$&$J_{\mathrm{K}_\mathrm{a}, \mathrm{K}_\mathrm{c}}=3_{2,2} - 2_{1,1}$&89.99&\tablefootmark{$\star$}1.49e-16 (22.9\%)\\
&p-H$_2$O&$\nu=0$&$J_{\mathrm{K}_\mathrm{a}, \mathrm{K}_\mathrm{c}}=7_{4,4} - 7_{3,5}$&90.05&8.09e-17 (24.0\%)\\
&o-H$_2$O&$\nu=0$&$J_{\mathrm{K}_\mathrm{a}, \mathrm{K}_\mathrm{c}}=6_{4,3} - 6_{3,4}$&92.81&\tablefootmark{b}1.20e-16 (21.8\%)\\
&p-H$_2$O&$\nu=0$&$J_{\mathrm{K}_\mathrm{a}, \mathrm{K}_\mathrm{c}}=7_{3,5} - 7_{2,6}$&93.38&9.90e-17 (22.4\%)\\
&o-H$_2$O&$\nu=0$&$J_{\mathrm{K}_\mathrm{a}, \mathrm{K}_\mathrm{c}}=6_{5,2} - 7_{2,5}$&94.17&\tablefootmark{$\star$}1.07e-16 (23.7\%)\\
&o-H$_2$O&$\nu=0$&$J_{\mathrm{K}_\mathrm{a}, \mathrm{K}_\mathrm{c}}=5_{4,2} - 5_{3,3}$&94.21&\\
&o-H$_2$O&$\nu_3=1$&$J_{\mathrm{K}_\mathrm{a}, \mathrm{K}_\mathrm{c}}=7_{4,4} - 7_{3,5}$&94.61&\tablefootmark{$\star$}1.10e-16 (23.4\%)\\
&o-H$_2$O&$\nu=0$&$J_{\mathrm{K}_\mathrm{a}, \mathrm{K}_\mathrm{c}}=6_{2,5} - 6_{1,6}$&94.64&\\
&o-H$_2$O&$\nu=0$&$J_{\mathrm{K}_\mathrm{a}, \mathrm{K}_\mathrm{c}}=4_{4,1} - 4_{3,2}$&94.71&1.40e-16 (22.5\%)\\
&p-H$_2$O&$\nu_2=1$&$J_{\mathrm{K}_\mathrm{a}, \mathrm{K}_\mathrm{c}}=5_{1,5} - 4_{0,4}$&94.90&4.91e-17 (32.5\%)\\*
&o-H$_2$O&$\nu=0$&$J_{\mathrm{K}_\mathrm{a}, \mathrm{K}_\mathrm{c}}=9_{4,5} - 8_{5,4}$&95.18&2.26e-17 (40.0\%)\\*
&p-H$_2$O&$\nu=0$&$J_{\mathrm{K}_\mathrm{a}, \mathrm{K}_\mathrm{c}}=5_{1,5} - 4_{0,4}$&95.63&8.59e-17 (28.8\%)\\
&p-H$_2$O&$\nu=0$&$J_{\mathrm{K}_\mathrm{a}, \mathrm{K}_\mathrm{c}}=4_{4,0} - 4_{3,1}$&95.88&\tablefootmark{b}9.46e-17 (26.3\%)\\
&o-H$_2$O&$\nu=0$&$J_{\mathrm{K}_\mathrm{a}, \mathrm{K}_\mathrm{c}}=5_{4,1} - 5_{3,2}$&98.49&\tablefootmark{$\star$}6.85e-17 (51.4\%)\\
&OH&$\nu=0$&$^2\Pi_{1/2} J=5/2 - ^2\Pi_{1/2} J=3/2$&98.72-98.74&\tablefootmark{$\star$,a}1.54e-16 (29.6\%) \\
&p-H$_2$O&$\nu_2=1$&$J_{\mathrm{K}_\mathrm{a}, \mathrm{K}_\mathrm{c}}=8_{3,5} - 7_{4,4}$&98.73&  \\\hline
R1A&p-H$_2$O&$\nu=0$&$J_{\mathrm{K}_\mathrm{a}, \mathrm{K}_\mathrm{c}}=6_{4,2} - 6_{3,3}$&103.92&\tablefootmark{$\star$}1.70e-16 (23.4\%)\\
&p-H$_2$O&$\nu=0$&$J_{\mathrm{K}_\mathrm{a}, \mathrm{K}_\mathrm{c}}=6_{1,5} - 6_{0,6}$&103.94&\\
&o-H$_2$O&$\nu=0$&$J_{\mathrm{K}_\mathrm{a}, \mathrm{K}_\mathrm{c}}=6_{3,4} - 6_{2,5}$&104.09&1.27e-16 (26.4\%)\\
&o-H$_2$O&$\nu_3=1$&$J_{\mathrm{K}_\mathrm{a}, \mathrm{K}_\mathrm{c}}=2_{2,0} - 1_{1,1}$&104.81&4.15e-17 (39.1\%)\\
&o-H$_2$O&$\nu=0$&$J_{\mathrm{K}_\mathrm{a}, \mathrm{K}_\mathrm{c}}=2_{2,1} - 1_{1,0}$&108.07&\tablefootmark{b}1.25e-16 (21.4\%)\\
&p-H$_2$O&$\nu=0$&$J_{\mathrm{K}_\mathrm{a}, \mathrm{K}_\mathrm{c}}=5_{2,4} - 5_{1,5}$&111.63&9.51e-17 (24.1\%)\\
&o-H$_2$O&$\nu=0$&$J_{\mathrm{K}_\mathrm{a}, \mathrm{K}_\mathrm{c}}=7_{4,3} - 7_{3,4}$&112.51&3.53e-17 (33.9\%)\\
&o-H$_2$O&$\nu_2=1$&$J_{\mathrm{K}_\mathrm{a}, \mathrm{K}_\mathrm{c}}=6_{4,3} - 7_{1,6}$&112.80&\tablefootmark{$\star$}3.20e-17 (39.4\%)\\
&o-H$_2$O&$\nu=0$&$J_{\mathrm{K}_\mathrm{a}, \mathrm{K}_\mathrm{c}}=4_{4,1} - 5_{1,4}$&112.80&\\
&o-H$_2$O&$\nu_3=1$&$J_{\mathrm{K}_\mathrm{a}, \mathrm{K}_\mathrm{c}}=5_{2,4} - 5_{1,5}$&112.89&\\
&o-H$_2$O&$\nu=0$&$J_{\mathrm{K}_\mathrm{a}, \mathrm{K}_\mathrm{c}}=4_{1,4} - 3_{0,3}$&113.54&9.06e-17 (23.7\%)\\
&p-H$_2$O&$\nu=0$&$J_{\mathrm{K}_\mathrm{a}, \mathrm{K}_\mathrm{c}}=5_{3,3} - 5_{2,4}$&113.95&1.24e-16 (21.9\%)\\
&o-H$_2$O&$\nu=0$&$J_{\mathrm{K}_\mathrm{a}, \mathrm{K}_\mathrm{c}}=4_{3,2} - 4_{2,3}$&121.72&\tablefootmark{b}9.60e-17 (21.2\%)\\
&p-H$_2$O&$\nu=0$&$J_{\mathrm{K}_\mathrm{a}, \mathrm{K}_\mathrm{c}}=8_{4,4} - 8_{3,5}$&122.52&2.63e-17 (34.0\%)\\
&o-H$_2$O&$\nu=0$&$J_{\mathrm{K}_\mathrm{a}, \mathrm{K}_\mathrm{c}}=9_{3,6} - 9_{2,7}$&123.46&3.11e-17 (30.1\%)\\
&o-H$_2$O&$\nu_2=1$&$J_{\mathrm{K}_\mathrm{a}, \mathrm{K}_\mathrm{c}}=5_{1,4} - 5_{0,5}$&124.85&\tablefootmark{$\star$}5.90e-17 (24.5\%)\\
&p-H$_2$O&$\nu=0$&$J_{\mathrm{K}_\mathrm{a}, \mathrm{K}_\mathrm{c}}=4_{0,4} - 3_{1,3}$&125.35&7.52e-17 (22.4\%)\\
&p-H$_2$O&$\nu=0$&$J_{\mathrm{K}_\mathrm{a}, \mathrm{K}_\mathrm{c}}=3_{3,1} - 3_{2,2}$&126.71&\tablefootmark{b}8.53e-17 (21.6\%)\\
&o-H$_2$O&$\nu=0$&$J_{\mathrm{K}_\mathrm{a}, \mathrm{K}_\mathrm{c}}=7_{2,5} - 7_{1,6}$&127.88&\tablefootmark{$\star$}6.84e-17 (26.4\%)\\
&o-H$_2$O&$\nu=0$&$J_{\mathrm{K}_\mathrm{a}, \mathrm{K}_\mathrm{c}}=9_{4,5} - 9_{3,6}$&129.34&4.27e-17 (24.0\%)\\
&o-H$_2$O&$\nu=0$&$J_{\mathrm{K}_\mathrm{a}, \mathrm{K}_\mathrm{c}}=4_{2,3} - 4_{1,4}$&132.41&8.39e-17 (21.3\%)\\
&o-H$_2$O&$\nu=0$&$J_{\mathrm{K}_\mathrm{a}, \mathrm{K}_\mathrm{c}}=5_{1,4} - 5_{0,5}$&134.94&\tablefootmark{b}7.45e-17 (22.2\%)\\
&o-H$_2$O&$\nu=0$&$J_{\mathrm{K}_\mathrm{a}, \mathrm{K}_\mathrm{c}}=3_{3,0} - 3_{2,1}$&136.50&8.01e-17 (21.7\%)\\
&p-H$_2$O&$\nu=0$&$J_{\mathrm{K}_\mathrm{a}, \mathrm{K}_\mathrm{c}}=3_{1,3} - 2_{0,2}$&138.53&\tablefootmark{$\star$}7.59e-17 (21.7\%)\\
&p-H$_2$O&$\nu=0$&$J_{\mathrm{K}_\mathrm{a}, \mathrm{K}_\mathrm{c}}=8_{4,4} - 7_{5,3}$&138.64&\\
&p-H$_2$O&$\nu_2=1$&$J_{\mathrm{K}_\mathrm{a}, \mathrm{K}_\mathrm{c}}=6_{3,3} - 6_{2,4}$&140.06&\tablefootmark{$\star$}2.99e-17 (31.7\%)\\
&p-H$_2$O&$\nu=0$&$J_{\mathrm{K}_\mathrm{a}, \mathrm{K}_\mathrm{c}}=4_{1,3} - 3_{2,2}$&144.52&\tablefootmark{b}4.63e-17 (23.5\%)\\\hline
R1B&p-H$_2$O&$\nu=0$&$J_{\mathrm{K}_\mathrm{a}, \mathrm{K}_\mathrm{c}}=4_{1,3} - 3_{2,2}$&144.52&\tablefootmark{b}4.80e-17 (23.7\%)\\
&p-H$_2$O&$\nu=0$&$J_{\mathrm{K}_\mathrm{a}, \mathrm{K}_\mathrm{c}}=4_{3,1} - 4_{2,2}$&146.92&6.72e-17 (22.0\%)\\
&o-H$_2$O&$\nu_3=1$&$J_{\mathrm{K}_\mathrm{a}, \mathrm{K}_\mathrm{c}}=8_{3,5} - 8_{2,6}$&148.64&\tablefootmark{$\star$}5.12e-17 (25.3\%)\\
&o-H$_2$O&$\nu=0$&$J_{\mathrm{K}_\mathrm{a}, \mathrm{K}_\mathrm{c}}=8_{3,5} - 8_{2,6}$&148.71&\\
&o-H$_2$O&$\nu=0$&$J_{\mathrm{K}_\mathrm{a}, \mathrm{K}_\mathrm{c}}=5_{4,2} - 6_{1,5}$&148.79&\\
&o-H$_2$O&$\nu_2=1$&$J_{\mathrm{K}_\mathrm{a}, \mathrm{K}_\mathrm{c}}=2_{2,1} - 2_{1,2}$&153.27&1.51e-17 (33.8\%)\\
&p-H$_2$O&$\nu_2=1$&$J_{\mathrm{K}_\mathrm{a}, \mathrm{K}_\mathrm{c}}=6_{2,4} - 6_{1,5}$&154.02&1.01e-17 (45.7\%)\\
&p-H$_2$O&$\nu=0$&$J_{\mathrm{K}_\mathrm{a}, \mathrm{K}_\mathrm{c}}=3_{2,2} - 3_{1,3}$&156.19&\tablefootmark{$\star$}1.28e-16 (20.7\%)\\
&p-H$_2$O&$\nu=0$&$J_{\mathrm{K}_\mathrm{a}, \mathrm{K}_\mathrm{c}}=5_{2,3} - 4_{3,2}$&156.27&\\
&o-H$_2$O&$\nu=0$&$J_{\mathrm{K}_\mathrm{a}, \mathrm{K}_\mathrm{c}}=5_{3,2} - 5_{2,3}$&160.51&\tablefootmark{b}4.60e-17 (23.6\%)\\
&OH&$\nu=0$&$^2\Pi_{1/2} J=3/2 - ^2\Pi_{1/2} J=1/2$&163.12-163.4&\tablefootmark{a}1.47e-16 (30.5\%) \\
&o-H$_2$O&$\nu=0$&$J_{\mathrm{K}_\mathrm{a}, \mathrm{K}_\mathrm{c}}=7_{3,4} - 7_{2,5}$&166.81&\tablefootmark{$\star$}3.77e-17 (31.7\%)\\
&o-H$_2$O&$\nu_3=1$&$J_{\mathrm{K}_\mathrm{a}, \mathrm{K}_\mathrm{c}}=6_{2,4} - 6_{1,5}$&166.83&\\
&p-H$_2$O&$\nu=0$&$J_{\mathrm{K}_\mathrm{a}, \mathrm{K}_\mathrm{c}}=6_{2,4} - 6_{1,5}$&167.03&4.11e-17 (27.9\%)\\
&p-H$_2$O&$\nu=0$&$J_{\mathrm{K}_\mathrm{a}, \mathrm{K}_\mathrm{c}}=6_{3,3} - 6_{2,4}$&170.14&\tablefootmark{$\star$}4.50e-17 (35.1\%)\\
&p-H$_2$O&$\nu=0$&$J_{\mathrm{K}_\mathrm{a}, \mathrm{K}_\mathrm{c}}=5_{3,3} - 6_{0,6}$&174.61&\tablefootmark{$\star$}8.25e-17 (21.9\%)\\
&p-H$_2$O&$\nu=0$&$J_{\mathrm{K}_\mathrm{a}, \mathrm{K}_\mathrm{c}}=3_{0,3} - 2_{1,2}$&174.63&\\
&p-H$_2$O&$\nu_3=1$&$J_{\mathrm{K}_\mathrm{a}, \mathrm{K}_\mathrm{c}}=3_{0,3} - 2_{1,2}$&174.66&\\
&o-H$_2$O&$\nu=0$&$J_{\mathrm{K}_\mathrm{a}, \mathrm{K}_\mathrm{c}}=4_{3,2} - 5_{0,5}$&174.92&1.49e-17 (43.1\%)\\
&o-H$_2$O&$\nu=0$&$J_{\mathrm{K}_\mathrm{a}, \mathrm{K}_\mathrm{c}}=2_{1,2} - 1_{0,1}$&179.53&\tablefootmark{b}4.93e-17 (24.1\%)\\
&o-H$_2$O&$\nu=0$&$J_{\mathrm{K}_\mathrm{a}, \mathrm{K}_\mathrm{c}}=2_{2,1} - 2_{1,2}$&180.49&\tablefootmark{b}5.87e-17 (26.7\%)\\
\end{longtable}
\end{longtab}
\twocolumn
   	\end{document}